\documentclass[10pt,fleqn,twocolumn]{article}
\usepackage[utf8]{inputenc}
\usepackage[english]{babel}
\pdfoutput=1
\usepackage{amsmath,amsfonts,amsthm,mathrsfs}
\usepackage{euscript}
\usepackage[full]{textcomp}
\usepackage[protrusion=true,expansion=true]{microtype}	
\usepackage{graphicx,color}
\usepackage[a4paper,includeheadfoot,left=1.3cm,right=1.3cm,top=1cm,head=0.5cm,bottom=1.7cm,foot=0.7cm]{geometry}
\usepackage{booktabs}
\usepackage{enumitem}
\usepackage[version=4]{mhchem}
\mhchemoptions{minus-sidebearing-left=0.5em,minus-sidebearing-right=0.5em}
\usepackage{flushend, balance} 

\usepackage[T1]{fontenc}
\usepackage[p,osf]{cochineal} 
\usepackage[varqu,varl,var0]{inconsolata}
\usepackage[scale=.95,type1]{cabin}
\usepackage[cochineal,vvarbb]{newtxmath}
\usepackage[scr=boondoxo]{mathalfa}

\usepackage{siunitx}
\DeclareSIUnit\gmol{g\text{-}mol}
\DeclareSIUnit\kgmol{kg\text{-}mol}
\DeclareSIUnit\lbmol{lb\text{-}mol}
\DeclareSIUnit\molar{\mole\per\cubic\deci\metre}
\DeclareSIUnit\Molar{M}
\DeclareSIUnit\torr{torr}
\DeclareSIUnit\micron{\micro\metre}
\DeclareSIUnit\mrad{\milli\rad}
\DeclareSIUnit\gauss{G}
\DeclareSIUnit\rpm{rpm}
\DeclareSIUnit\inch{in}
\DeclareSIUnit\watt{W}
\DeclareSIUnit\ppm{ppm}
\DeclareSIUnit\sccm{sccm}

\usepackage[compress, numbers, super]{natbib}
\setcitestyle{round} 

\usepackage[font=small,labelfont={color=black,bf},textfont={color=black},format=plain,indention=0cm,justification=centerlast,skip=0.1cm]{caption,subfig}
\makeatletter
\renewcommand{\fnum@figure}{\textbf{\small\mbox{Fig.~\thefigure}}}
\renewcommand{\fnum@table}{\textbf{\small\mbox{Table~\thetable}}}
\makeatother

\DeclareRobustCommand{\uppartial}{\text{\rotatebox[origin=t]{19}{\scalebox{0.99}[1]{$\partial$}}}\hspace{-1pt}}

\newcommand{\erf}[1]{\mathrm{erf}{#1}}
\newcommand{\erfc}[1]{\mathrm{erfc}{#1}}

\newcommand{\mytitle}{\textbf{Gas diffusion in nanoporous thin films}}
\newcommand{\linefindoc}{\color{black}
\vspace{0.5cm}
\centering\rule{0.7\linewidth}{1.2pt}\\%
\vspace{-0.39cm}
\rule{0.5\linewidth}{1.2pt}\\%
\vspace{-0.39cm}
\rule{0.3\linewidth}{1.2pt}%
\vspace{-0.5cm}}
\DeclareRobustCommand{\uppartial}{\text{\rotatebox[origin=t]{15}{\scalebox{0.95}[1]{$\partial$}}}\hspace{-1pt}} 
\newcommand{\hyphen}[1]{---{#1}---} 

\newenvironment{rcases}
  {\left.\begin{aligned}}
  {\end{aligned}\quad\right\rbrace}

\usepackage[colorlinks=true]{hyperref}
\hypersetup{
pdffitwindow=true,
pdftitle={\mytitle},
pdfauthor={L. N. Acquaroli},
pdfsubject={},
pdfcreator={L.N. Acquaroli},
pdfproducer={L.N. Acquaroli},
pdfkeywords={},
colorlinks={true},
linkcolor={NeonBlue},
citecolor={Cinnabar},
filecolor={Cinnabar},
urlcolor={NeonBlue}
}
\usepackage{url} 

\definecolor{NeonBlue}{rgb}{0.11,0.22,0.73} 
\definecolor{Cinnabar}{rgb}{0.8078,0.0863,0.1255}

\usepackage{titlesec}
\titlelabel{\thetitle. }
\titleformat*{\section}{\centering\large\bfseries}
\titleformat*{\subsection}{\centering\bfseries}

\usepackage{titling}	

\pretitle{\vspace{-40pt} \begin{center} \vspace{5pt} \fontsize{14}{14} \bfseries \color{black} \selectfont }
\title{\mytitle}
\posttitle{\par\end{center}\vspace{-.15cm}}
\preauthor{\vspace{-13pt} \begin{center}
\bigskip \color{black}}
\author{\textbf{Leandro N. Acquaroli}}	
\postauthor{\small \color{black}
\medskip

{Department of Engineering Physics, Ecole Polytechnique Montreal

P.O. Box 6079, Station Centre-Ville, Montreal (QC) H3C 3A7, Canada} 
\vspace{-0.4cm}
\date{\small\today}
\par\end{center}}

\usepackage{fancyhdr}	
\usepackage{lastpage}
\renewcommand{\headrulewidth}{0.0pt}
\renewcommand{\footrulewidth}{0.0pt}

\usepackage{abstract}
\usepackage{indentfirst}

\renewcommand{\thetable}{\Roman{table}}

\begin{document}

\setlength{\belowdisplayskip}{5pt}\setlength{\belowdisplayshortskip}{5pt}
\setlength{\abovedisplayskip}{5pt}\setlength{\abovedisplayshortskip}{5pt}

\columnsep 0.6cm

\renewcommand{\abstractname}{}
\twocolumn[
\maketitle
\vspace{-1.7cm}
\begin{onecolabstract}
We analyze the Fick's diffusion of a gas inside porous nanomaterials through the one-dimensional diffusion equation in nanopores for various cases of boundary conditions for homogeneous and non-homogeneous problems. We study the diffusion problems, starting without adsorption of the gas inside the pores, to more complex situations with surface adsorption in the pore walls and at the pore tips. Different methods of solution are reviewed depending on the problem, such as similarity transformation, Laplace transform, separation of variables, Danckwerts method and the Green's functions technique. The recovery step when the diffusion process stops and reached the steady-state is presented as well for the different problems.
\end{onecolabstract}
\vspace{1cm}
]

\fancypagestyle{plain}{%
\fancyhf{} 
\fancyfoot[R]{\footnotesize \thepage\ of \pageref{LastPage}} 
\renewcommand{\headrulewidth}{0pt}
\renewcommand{\footrulewidth}{0pt}}

\section{Introduction}

In the last decades, research on sensors fabricated with nanomaterials increased due to their broad range of applications from optoelectronics to sensors~\cite{bisi1, theiss1, monsouri1}.

Porous silicon \hyphen{PS} is a nanomaterial obtained by electrochemical anodization of crystalline silicon \hyphen{c-Si} wafers in hydrofluoric acid \hyphen{HF} solutions containing a surfactant such as ethanol \hyphen{EtOH}. Under proper preparation conditions, a porous network grows inside the c-Si wafer with pores sizes varying from \SI{2}{\nano\meter} up to \SI{10}{\micro\meter}, and surface areas up to \SI{800}{\square\meter\per\gram}. Due to these properties, its fast preparation and its diverse and tuneable optical, electrical and surface-chemical properties predestined PS for sensor and biosensor applications~\cite{adfm.200500218, adfm.200500899, ACQUAROLI2010189, 10408430903245369, Haidary, C4RA04184D, doi:10.1002/adtp.201800095}.

\begin{figure}[t]
   \begin{center}
       \includegraphics[scale=0.4]{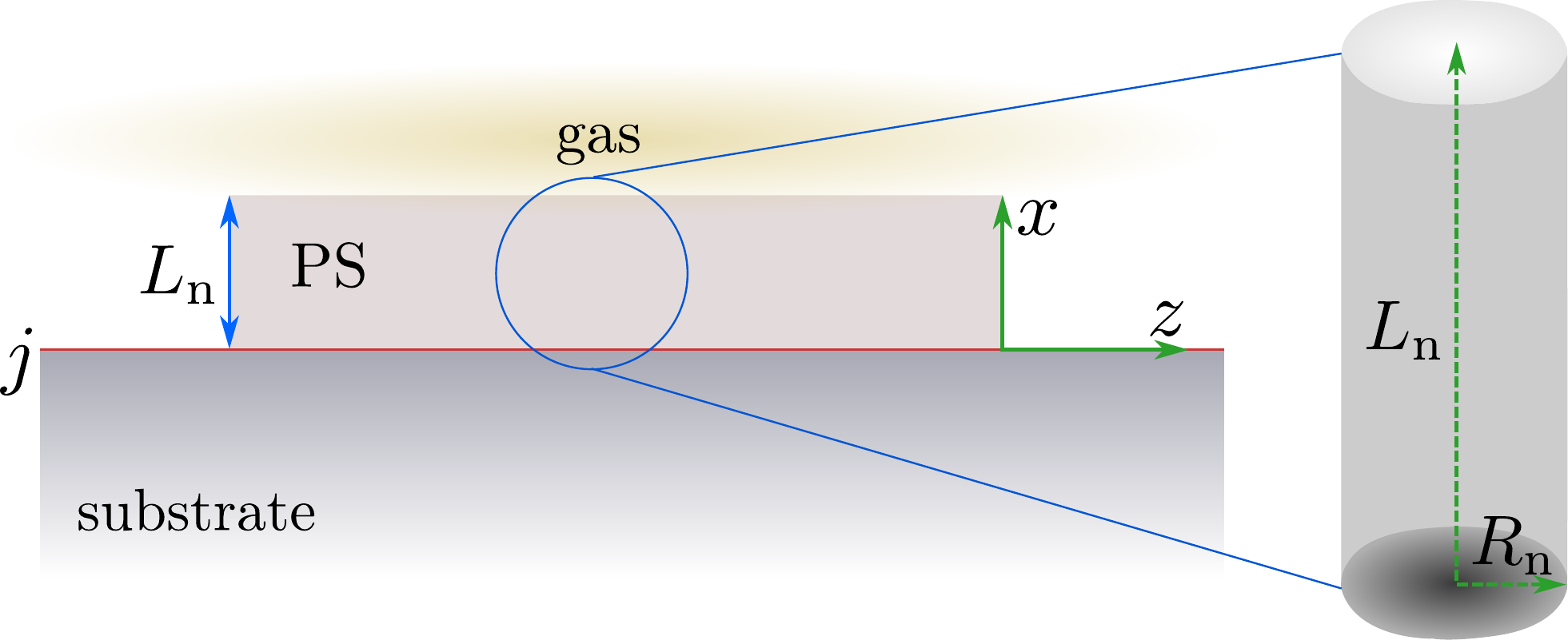}
       \vspace{0.2cm}
       \caption{Scheme of a nanoporous thin film surrounded by gas at the top and sitting on a solid substrate. The detail on the right shows an idealized cylindrical geometry used to analyze transient axisymmetric diffusion in a nanopore.\hfill{ }}
       \label{f.fig1}
   \end{center}
 \end{figure}

We present an analysis of the different approaches to the one-dimensional diffusion equation, considering nanoscale mass transport inside a porous material. 
Consider a porous thin film on a substrate with an idealized cylindrical close-end pore \hyphen{so that there is no net hydrodynamic flow into the pore, thus, neglecting advective transport} with length $L_{\text{n}}$ and radius $R_{\text{n}}$ \hyphen{Fig.~\ref{f.fig1}}. The transport always occurs in gas phase by neglecting Kelvin condensation in the nanopores. Adsorption and desorption time scales vary depending upon substrate surface and analyte species, but they typically fall in the range $\approx$\SIrange{e-6}{e-3}{\second}, compared to the diffusion times in the \SI{e-11}{\second}. Considering the length and time scales above, the justifications for the use of the continuum assumption are not applicable. The requirement that the characteristic system length scale is large when compared to the characteristic molecular length scale is equivalent to the requirement that the Knudsen number, $K_{\text{n}}\equiv \mu\sqrt{RT/M}/(P R_{\text{n}})$, be small. For instance, in PS nanopores, $K_{\text{n}}\approx 5$. Thus, it would appear that the continuum assumption should not be applied and that the governing equation for mass transport should be the more general Boltzmann transport equation. However, it has been demonstrated that a Fickian-like continuum model can still be used to describe mass transport in gasses at moderate Knudsen numbers provided the diffusion coefficient is appropriately modified to account for rarefaction effects~\cite{Kottke2009}.

We present the diffusion problems, from simple without adsorption to more complex situations with surface adsorption in the pore walls and at the pore tips. We review different methods of solution depending on the problem and the boundary conditions, such as similarity transformation, Laplace transform, separation of variables, Danckwerts method and the Green's functions technique. We study as well the recovery step, when the diffusion process stops and reached the steady-state, for the different problems.

\section{Diffusion problem at the nanoscale}
A simplified version of the problem is reached assuming that the difference between the concentration at the pore wall and a cross-sectional area weighted average concentration are negligible. At least for short times, the fraction of active sites per unit area filled with the adsorbed species is negligible, and only adsorption \hyphen{no desorption} occurs at an appreciable rate. Thus, removing the radial coordinate, the general dimensionless governing problem results as follow:
\begin{subequations}\label{eq.1}
\begin{align}
    \text{(DE):}\quad & \uppartial_t c^* = \uppartial_{xx} c^* - \beta \alpha^2 c^*,\,\,\,\, 0<x^* <1,\,\, t^* >0, \label{eq.1a}\\
    \text{(BC-1):}\quad & c^*(0, t^*) = 1,\,\,\,\, t^* >0, \label{eq.1b}\\
    \text{(BC-2):}\quad & \uppartial_x c^*(1, t^*) + \gamma \alpha^2 c^*(1, t^*)=0,\,\,\,\, t^* >0, \label{eq.1c}\\
    \text{(IC):}\quad & c^*(x^*, 0) = 0,\,\,\,\, 0<x^* <1, \label{eq.1d}
\end{align}
\end{subequations}

\noindent
where $c^*(x^*, t^*)=c(x,t)/c_0$ is the concentration normalized to the initial $c_0$, $x^* = x/L_{\text{n}}$ is the dimensionless space coordinate, $t^* = t\, D_{\text{Kn}}/L_{\text{n}}^2$ is the dimensionless time, $D_{\text{Kn}} = R_{\text{n}} \sqrt{RT/M}$ is the Knudsen diffusion, with $R$ is the universal gas constant, $T$ is the temperature and $M$ is the mole average molecular weight of the gas mixture. The parameter $\alpha = k_{\text{a}} L_{\text{n}} N/D_{\text{Kn}}$ where $k_{\text{a}}$ is the rate coefficient for adsorption, and $N$ the number of active sites per unit area of surface~\cite{Kottke2009}. The parameters $\beta$ and $\gamma$ serve to describe the general problem and will take values 0 or 1 depending on the different problems and boundary conditions considered.

 \section{Recovery process from diffusion}
  For sensing purposes, is interesting to determine the mathematical solution to the problem of diffusion, once the steady-state is reached and the point of diffusion is off. The steady-state is calculated taking the limit of the solutions of the diffusion problems discussed before when the time tends to infinity. Switching off the source of diffusion is done by setting the Dirichlet BC-1 \eqref{eq.1b} to zero, giving the general problem for the normalized concentration $u^*(x^*,t^*)$ as follow:
 \begin{subequations}\label{eq.62}
 \begin{align}
     \text{(DE):}\quad & \uppartial_t u^* = \uppartial_{xx} u^*, \label{eq.62a}\\
     \text{(BC-1):}\quad & u^*(0, t^*) = 0, \label{eq.62b}\\
     \text{(BC-2):}\quad & \uppartial_x u^*(1, t^*) = 0, \label{eq.62c}\\
     \text{(IC):}\quad & u^*(x^*, 0) = c^*(x^*,\infty). \label{eq.62d}
 \end{align}
 \end{subequations}

\section{Solutions to the diffusion problem without adsorption}
We will tackle different approaches to solve the system \eqref{eq.1} depending on the conditions established. First, we will consider the problem without adsorption, \text{i.e.}, setting $\beta=\gamma=0$ in \eqref{eq.1} and solving the problem simplest diffusion equation with Dirichlet and Neumann boundary conditions using three different approaches.

\subsection{Similarity transformation}
Similarity transformations to PDEs are solutions which relies on treating independent variables in groups rather than separately. Setting $\beta=\gamma=0$ in problem~\eqref{eq.1} gives:
\begin{subequations}\label{eq.2}
    \begin{align}
        \text{(DE):}\quad & \uppartial_t c^* = \uppartial_{xx} c^*, \label{eq.2a}\\
        \text{(BC-1):}\quad & c^*(0, t^*) = 1, \label{eq.2b}\\
        \text{(BC-2):}\quad & \uppartial_x c^*(1, t^*) = 0, \label{eq.2c}\\
        \text{(IC):}\quad & c^*(x^*, 0) = 0, \label{eq.2d}
    \end{align}
\end{subequations}

\noindent
We seek a solution of the form\cite{Crank1975}
\begin{equation}\label{eq.3}
    c^* = (t^*)^r\, g(\eta^*),\quad \eta^* = \frac{x^*}{\sqrt{t^*}},
\end{equation}

\noindent
where $r$ is chosen arbitrarily to satisfy the BCs. Then, we plug~\eqref{eq.3} into~\eqref{eq.2a}:
\begin{subequations}\label{eq.4}
    \begin{align}
        \uppartial_t c^* &= (t^*)^{r-1}\, rg - (t^*)^{r-1}\, \frac{\eta^*\, g'}{2}, \label{eq.4a}\\
        \uppartial_{xx} c^* &= (t^*)^{r-1}\, g''.
    \end{align}
\end{subequations}

\noindent
We set $r=1/2$ to satisfy the BCs in the final transformed ODE:
\begin{subequations}\label{eq.5}
    \begin{align}
        \text{(DE):}\quad & g'' + \frac{\eta^*}{2} g' - \frac{1}{2}g = 0, \label{eq.5a}\\
        \text{(BC-1):}\quad & g(\eta^*\to 0) \to 1, \label{eq.5b}\\
        \text{(BC-2):}\quad & g'(\eta^*\to\infty) \to 0. \label{eq.5c}
    \end{align}
\end{subequations}

\noindent
Using the transformation $g(\eta^*)=\eta^* f(\eta^*)$ in \eqref{eq.5a} gives the differential equation: $\eta^* f''+[2+(\eta^*)^2/2] f'=0$. Solving for $f$ and applying the anti-transformation, we get:
\begin{equation}\label{eq.6}
    g(\eta^*) = \frac{1}{2}\left[ 2\, \exp\!{(-(\eta^*)^2 /4)} - \sqrt{\pi}\,\eta^* + \eta^*\, \erf{(\eta^*/2)} \right].
\end{equation}

\noindent
Thus, combining Eqs.~\eqref{eq.3} and \eqref{eq.6}, the solution of \eqref{eq.2} results as follow:
\begin{equation}\label{eq.7}
    c^*(x^*,t^*) = \sqrt{t^*}\,\exp\!{\left(-\frac{x^*}{4t^*}\right)} - \frac{\sqrt{\pi}}{2} x^* \erfc{\left(\frac{x^*}{2 \sqrt{t^*}}\right)}.
\end{equation}

\noindent
Integration of \eqref{eq.7} over $x^*$ yields the time solution of the concentration:
\begin{align}
    c^*(t^*) &= \int_0^1 c^*(x^*,t^*)\, \text{d}x^* \label{eq.8}\\
    &= \frac{1}{2}\, \sqrt{t^*}\, \exp\!{\left(-\frac{1}{4t^*}\right)} + \frac{\sqrt{\pi}}{4} \left[ (1 + 2t^*)\,\erf{\left(\frac{1}{2 \sqrt{t^*}}\right)} - 1 \right]. \label{eq.9}
\end{align}

\begin{figure}[t]
   \begin{center}
       \hspace{-0.1cm}\includegraphics[scale=0.425]{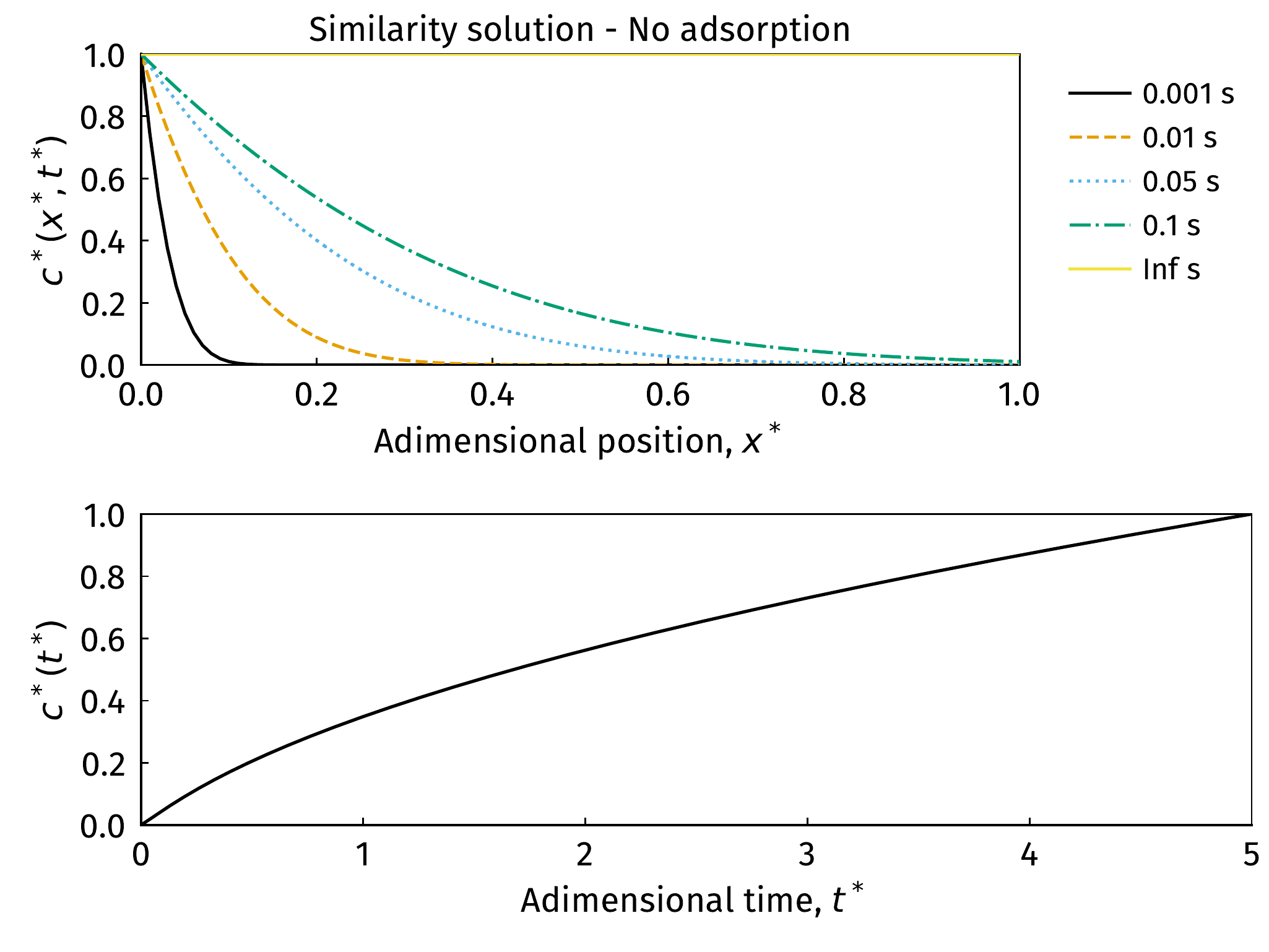}
       \caption{Concentration profiles of the similarity solution: (top) Eq.~\eqref{eq.7} with legends indicating different $t^*$; (bottom) Eq.~\eqref{eq.9}. \hfill{ }}
       \label{f.similarity_solution}
   \end{center}
 \end{figure}

Equation~\eqref{eq.7} shows a decay behavior which strongly depends on $t^*$, while the time profile of Eq.~\eqref{eq.9} builds up until it reaches the end of the pore, without clear saturation \hyphen{Fig.~\ref{f.similarity_solution}}.

\subsection{Laplace transform}
The method of Laplace transform is widely used in the solution of time-dependent diffusion problems because the partial derivative with respect to the time variable is removed from the differential equation of diffusion and replaced with a parameter, $s$, in the transformed field. Thus, this technique to solve PDEs is relatively straightforward, however, the inversion of the transformed solution generally is rather involved unless the inversion is available in the standard Laplace transform tables\cite{Haberman2012, Ozisik2012}. In fact, to simplify the inverse of the Laplace transform in our problem, we change the limits of the BCs, as if we flipped the thin film in the spatial coordinate. By setting $\beta=\gamma=0$ in \eqref{eq.1}, we get:
\begin{subequations}\label{eq.10}
    \begin{align}
        \text{(DE):}\quad & \uppartial_t c^* = \uppartial_{xx} c^*, \label{eq.10a}\\
        \text{(BC-1):}\quad & \uppartial_x c^*(0, t^*) = 0, \label{eq.10b}\\
        \text{(BC-2):}\quad & c^*(1, t^*) = 1, \label{eq.10c}\\
        \text{(IC):}\quad & c^*(x^*, 0) = 0, \label{eq.10d}
    \end{align}
\end{subequations}

We start by taking the Laplace transform, $\mathcal{L}$, of each term of \eqref{eq.10} as follow:
\begin{subequations}\label{eq.11}
    \begin{align}
        \mathcal{L}[\uppartial_{xx} c^*] &= \hat{c}''(x^*,s), \label{eq.11a}\\
        \mathcal{L}[\uppartial_t c^*] &= s\,\hat{c}(x^*,s) - c^*(x^*,0) = s\,\hat{c}(x^*,s), \label{eq.11b}\\
        \mathcal{L}[\uppartial_x c^*(0, t^*)] &= \mathcal{L}[0] = \hat{c}'(0, s) = 0, \label{eq.11c}\\
        \mathcal{L}[c^*(1, t^*)] &= \mathcal{L}[1] = \hat{c}(1, s) = \frac{1}{s},\label{eq.11d}
    \end{align}
\end{subequations}

\noindent
where the variable $\hat{c}=\hat{c}(x^*,s)$ represents the concentration in the transformed field, where $s$ is a parameter, not a variable. Then, the PDE of problem \eqref{eq.10} is transformed into the following ODE:
\begin{subequations}\label{eq.12}
    \begin{align}
        \text{(DE):}\quad & \hat{c}'' - s \hat{c} = 0, \label{eq.12a}\\
        \text{(BC-1):}\quad & \hat{c}'(0, s) = 0, \label{eq.12b}\\
        \text{(BC-2):}\quad & \hat{c}(1, s) = \frac{1}{s}, \label{eq.12c}
    \end{align}
\end{subequations}

\noindent
The general solution for problem \eqref{eq.12} has the form\cite{Ozisik2012}
\begin{subequations}\label{eq.13}
    \begin{align}
        \hat{c} &= A_1 \cosh(x^*\sqrt{s}) + A_2 \sinh(x^*\sqrt{s}),\label{eq.13a}\\
        \hat{c}' &= A_1 \sqrt{s} \sinh(x^*\sqrt{s}) + A_2 \sqrt{s} \cosh(x^*\sqrt{s}).\label{eq.13b}
    \end{align}
\end{subequations}

\noindent
Applying the BCs to the solution \eqref{eq.13}, gives $A_1 = [s\cosh(\sqrt{s})]^{-1}$ and $A_2 = 0$. Plugging these constants into \eqref{eq.13a}, we get the solution in the transformed field:
\begin{equation}\label{eq.14}
    \hat{c}(x^*,s) = \frac{\cosh(x^*\sqrt{s})}{s\cosh(\sqrt{s})}.
\end{equation}

\noindent
The inverse of the Laplace transform, $\mathcal{L}^{-1}$, of \eqref{eq.14}, is found in tables and gives the following solution to \eqref{eq.10}\cite{Ozisik2012}:
\begin{align}
    c^*(x^*,t^*) &= \mathcal{L}^{-1}[\hat{c}(x^*,s)] \label{eq.15} \\
    &= 1 + 2 \sum_{n=1}^{\infty} \frac{(-1)^n}{\lambda_n}\cos(\lambda_n x^*) \exp(-\lambda_n^2 t^*).\label{eq.16}
\end{align}

\noindent
where $\lambda_n=(n-1/2)\pi, n\in\mathbb{Z}^+$. The time profile is \hyphen{Eq.~\eqref{eq.8}}:
\begin{equation}\label{eq.17}
    c^*(t^*) = 1 - 2 \sum_{n=1}^{\infty} \frac{1}{\lambda_n^2} \exp(-\lambda_n^2 t^*).
\end{equation}

\begin{figure}[t]
   \begin{center}
       \hspace{-0.1cm}\includegraphics[scale=0.425]{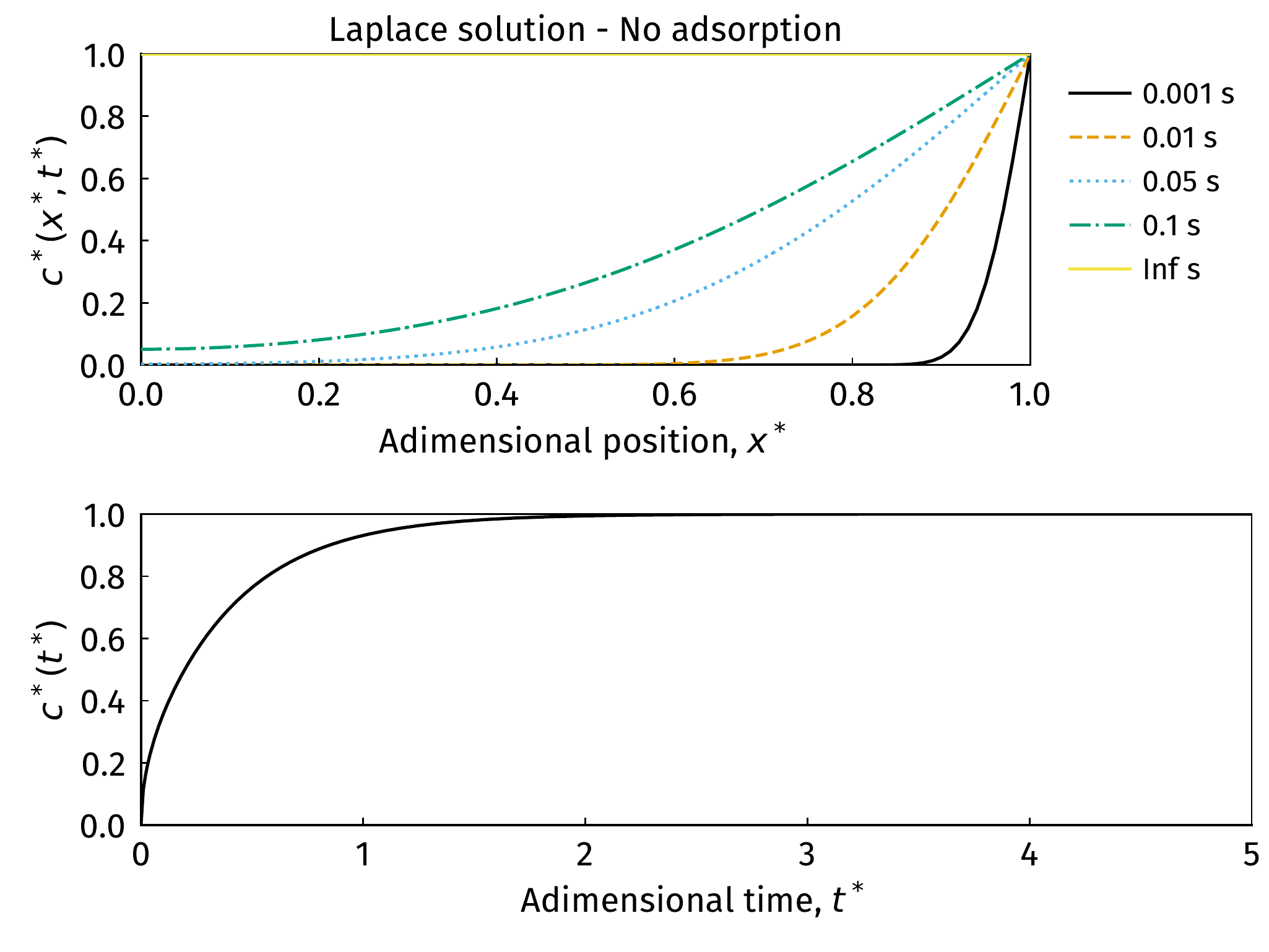}
       \caption{Concentration profiles using the Laplace transform solution: (top) Eq.~\eqref{eq.16} with legends indicating different $t^*$; (bottom) Eq.~\eqref{eq.17}. Note the inversion of the boundary conditions. \hfill{ }}
       \label{f.laplace_solution}
   \end{center}
 \end{figure}

Solutions obtained with Laplace transform are plotted in Fig.~\ref{f.laplace_solution}.

\subsection{Separation of variables}
The inverse of Laplace transform is useful when the anti-transformation is generally tabulated. Otherwise, it involves further complicated steps depending on the solution in the transformed field. The method of separation of variables presents as an alternative. To use this method, we split the problem \eqref{eq.2} into an equilibrium steady-state part, $v(x^*)$, plus a non-equilibrium displacement part, $w(x^*,t^*)$\cite{Haberman2012}. By doing this, we send the non-homogeneity in BC-1 to the IC. Thus,
\begin{equation}\label{eq.18}
    c^*(x^*,t^*) = v(x^*) + w(x^*,t^*),
\end{equation}

\noindent
The steady-state part problem is:
\begin{equation}\label{eq.19}
    \begin{rcases}
        \text{(DE):}\quad & v''(x^*) = 0\\
        \text{(BC-1):}\quad & v(0) = 1\\
        \text{(BC-2):}\quad & v'(1) = 0
    \end{rcases}
    \,\, v(x^*)=1.
\end{equation}

\noindent
The non-equilibrium problem is built from expression \eqref{eq.18}, $w^*(x^*,t^*) = c(x^*,t^*) - v(x^*)$, with homogenous BCs as follow:
\begin{subequations}\label{eq.20}
    \begin{align}
        \text{(DE):}\quad & \uppartial_t w = \uppartial_{xx} w, \label{eq.20a}\\
        \text{(BC-1):}\quad & w(0,t^*) = 0, \label{eq.20b}\\
        \text{(BC-2):}\quad & \uppartial_x w(1,t^*) = 0, \label{eq.20c}\\
        \text{(IC):}\quad & w(x^*,0) = c^*(x^*,0)-v(x^*)=-v(x^*), \label{eq.20d}
    \end{align}
\end{subequations}

\noindent
where $c^*(x^*,0)=0$ in the IC results from \eqref{eq.10d}. Replacing $w(x^*,t^*)=p(x^*)\,q(t^*)$ in \eqref{eq.20a} gives:
\begin{equation}\label{eq.21}
    \frac{p''}{p} = \frac{q'}{q} = -\lambda^2,
\end{equation}

\noindent
where $\lambda$ is an arbitrary separation constant. Last expression yields two ordinary differential equations for $x^*$ and $t^*$. The $x^*$-dependent part satisfies the eigenvalue problem with two homogeneous boundary conditions as follow:
\begin{subequations}\label{eq.22}
    \begin{align}
        \text{(DE):}\quad & p'' + \lambda^2 p = 0, \label{eq.22a}\\
        \text{(BC-1):}\quad & p(0) = 0, \label{eq.22b}\\
        \text{(BC-2):}\quad & p'(1) = 0. \label{eq.22c}
    \end{align}
\end{subequations}

\noindent
The general solution of \eqref{eq.22} is
\begin{subequations}\label{eq.23}
    \begin{align}
        p(x^*) &= A_1 \cos(\lambda x^*) + A_2 \sin(\lambda x^*),\label{eq.23a}\\
        p'(x^*) &= -A_1 \lambda \sin(\lambda x^*) + A_2 \lambda \cos(\lambda x^*).\label{eq.23b}
    \end{align}
\end{subequations}

\noindent
Placing the BCs we obtain the nontrivial solutions $A_1=0$ and $A_2\cos(\lambda)=0$. Since the cosine is a periodic function, the eigenvalue $\lambda$ must satisfy the solution for every positive odd half-integer of $\pi$, hence, the eigenvalues are given by
\begin{equation}\label{eq.24}
    \lambda_n = \left(n-\frac{1}{2}\right)\pi,\quad n\in\mathbb{Z}^+.
\end{equation}

\noindent
The eigenfunction corresponding to the eigenvalue $\lambda_n$ is
\begin{equation}\label{eq.25}
    p_n(x^*) = A_2 \sin(\lambda_n x^*).
\end{equation}

The time-dependent ODE equation that results from \eqref{eq.21}
\begin{equation}\label{eq.26}
    q'(t^*) + \lambda^2 q(t^*) = 0
\end{equation}

\noindent
has the following general solution:
\begin{equation}\label{eq.27}
    q_n(t^*) = q(0)\,\exp(-\lambda^2_n t^*).
\end{equation}

\noindent
Therefore, the product solution of \eqref{eq.20} is
\begin{equation}\label{eq.28}
    w_n(x^*,t^*) = B_n\,\sin(\lambda_n x^*)\,\exp(-\lambda^2_n t^*).
\end{equation}

\noindent
where $B_n=A_2\,q(0)$. The principle of superposition shows that $w_n$, with $n\in\mathbb{Z}^+$, are solutions of a linear homogeneous problem. It follows that any linear combination of these solutions is also a solution of the linear homogeneous equation \eqref{eq.20}. Thus, taking into account that $B_n$ could be different for each solution, we have
\begin{equation}\label{eq.29}
    w(x^*,t^*) = \sum_{n=1}^{\infty} B_n \sin(\lambda_n x^*)\,\exp(-\lambda^2_n t^*).
\end{equation}

\noindent
The initial condition is satisfied if
\begin{equation}\label{eq.30}
    w(x^*,0) = -v(x^*) = \sum_{n=1}^{\infty} B_n \sin(\lambda_n x^*).
\end{equation}

\noindent
We first multiply both sides of \eqref{eq.30} by $p_m(x^*)$ \hyphen{for a given integer $m$ of $n$}, and integrate over $x^*$:
\begin{equation}\label{eq.31}
    -\int_0^1 v(x^*) \sin(\lambda_m x^*)\, \text{d}x^*= \sum_{n=1}^{\infty} B_n \int_0^1 \sin(\lambda_n x^*) \sin(\lambda_m x^*)\, \text{d}x^*.
\end{equation}

\begin{figure}[t]
   \begin{center}
       \hspace{-0.1cm}\includegraphics[scale=0.425]{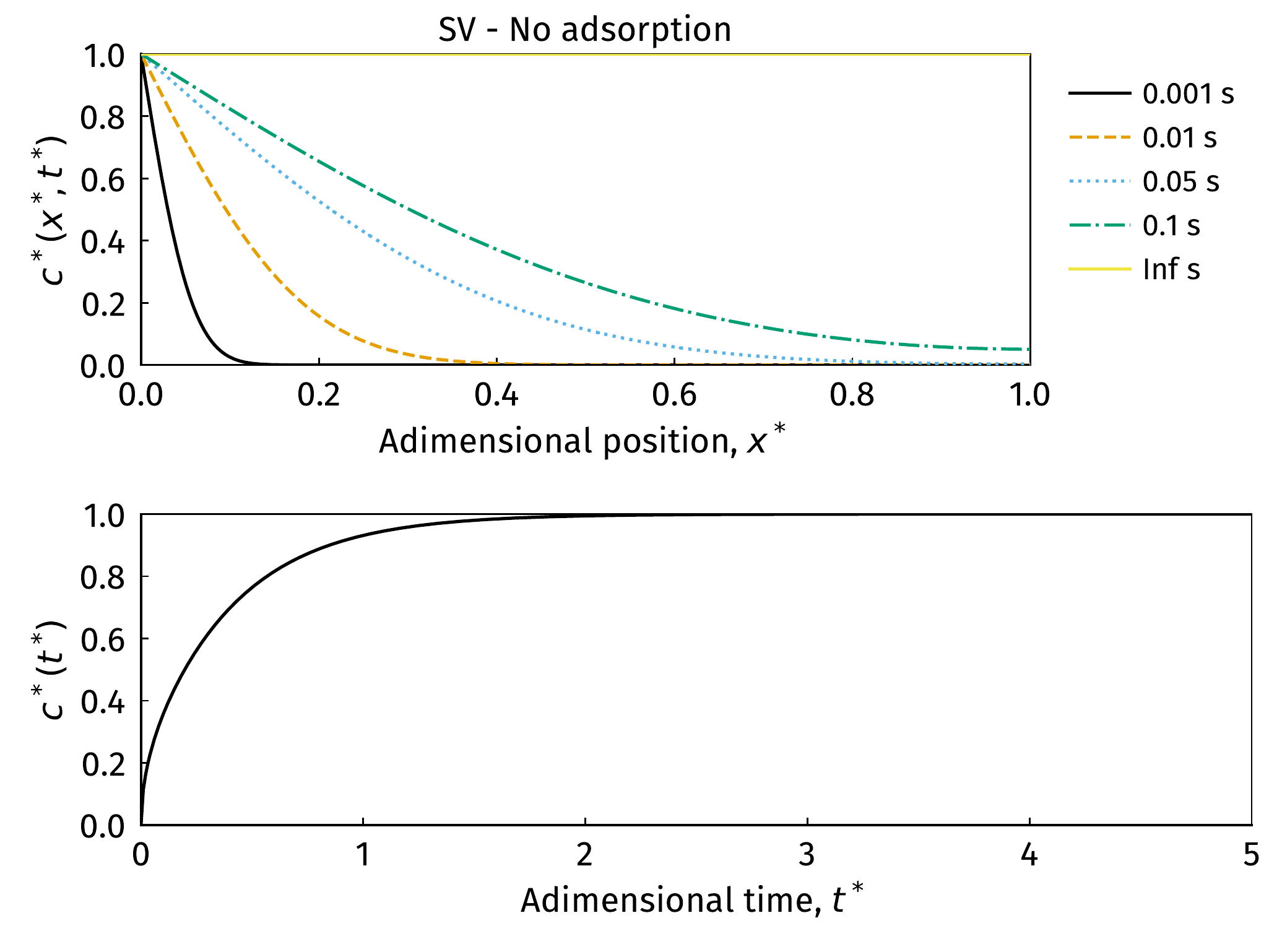}
       \caption{Concentration profiles using the separation of variables solution: (top) Eq.~\eqref{eq.35} with legends indicating different $t^*$; (bottom) Eq.~\eqref{eq.36}. \hfill{ }}
       \label{f.sv_solution_no_ads}
   \end{center}
\end{figure}

\noindent
The orthogonality property of the sine function implies that each term of the sum is zero whenever
$n \ne m$, then, the only term that appears on the right-hand side occurs when $m$ is replaced by $n$:
\begin{equation}\label{eq.32}
    -\int_0^1 v(x^*) \sin(\lambda_n x^*)\, \text{d}x^*= B_n \int_0^1 \sin^2(\lambda_n x^*)\, \text{d}x^*.
\end{equation}

\noindent
Solving for $B_n$ we obtain:
\begin{equation}\label{eq.33}
    B_n = -\frac{\int_0^1 v(x^*) \sin(\lambda_n x^*)\, \text{d}x^*}{\int_0^1 \sin^2(\lambda_n x^*)\, \text{d}x^*} = -2 \int_0^1 v(x^*) \sin(\lambda_n x^*)\, \text{d}x^* .
\end{equation}

\noindent
Replacing $v(x^*)=1$, results in
\begin{equation}\label{eq.34}
    B_n = - 2 \int_0^1 \sin(\lambda_n x^*)\, \text{d}x^* = -\frac{2}{\lambda_n}.
\end{equation}

\noindent
Combining Eqs.~\eqref{eq.18}, \eqref{eq.29} and \eqref{eq.34}, the final solution to the problem is:
\begin{equation}\label{eq.35}
    c^*(x^*,t^*) = 1 - \sum_{n=1}^{\infty} \frac{2}{\lambda_n} \sin(\lambda_n x^*)\exp(-\lambda^2_n t^*)
\end{equation}

\noindent
where the time-profile defined as
\begin{equation}\label{eq.36}
    c^*(t^*) = 1 - \sum_{n=1}^{\infty} \frac{2}{\lambda_n^2} \exp(-\lambda^2_n t^*).
\end{equation}

\noindent
and the eigenvalues given by \eqref{eq.24}. Plots of \eqref{eq.35} and \eqref{eq.36} are shown in Fig.~\ref{f.sv_solution_no_ads}.

\section{Solutions to the diffusion problem with adsorption}
Now we will consider two approaches to solve the system \eqref{eq.1} with adsorption. There are solutions to this problem in literature, for $\beta=1$ and $\gamma=0$, giving rise to the surface adsorption effect inside the pores' walls, except at the bottom of the pore. We review this problem and also the solution for the Robin \hyphen{mixed} boundary condition, $\gamma=1$, when the adsorption at the pore-tip is considered.

\subsection{Adsorption in the wall's surface of the pore}
Setting $\beta=1$ and $\gamma=0$ in \eqref{eq.1} becomes a reaction-diffusion problem with Dirichlet and Neumann boundary conditions as follow:
\begin{subequations}\label{eq.37}
\begin{align}
    \text{(DE):}\quad & \uppartial_t c^* = \uppartial_{xx} c^* - \alpha^2 c^*, \label{eq.37a}\\
    \text{(BC-1):}\quad & c^*(0, t^*) = 1, \label{eq.37b}\\
    \text{(BC-2):}\quad & \uppartial_{x} c^*(1, t^*) = 0, \label{eq.37c}\\
    \text{(IC):}\quad & c^*(x^*, 0) = 0. \label{eq.37d}
\end{align}
\end{subequations}

\noindent
In the literature, exists solutions to this problem using eigenfunction expasion\cite{Kottke2009}, superposition of solutions\cite{MATSUNAGA2003226} and Laplace transform\cite{SELVARAJ2014885}. Here, taking advantage to the constant nature of the BCs we solve this problem using the Danckwerts method\cite{Crank1975}, which consists in solving the homogeneous DE, keeping the BCs and IC,
\begin{subequations}\label{eq.38}
\begin{align}
    \text{(DE):}\quad & \uppartial_t\hat{c} = \uppartial_{xx}\hat{c}, \label{eq.38a}\\
    \text{(BC-1):}\quad & \hat{c}(0, t^*) = 1, \label{eq.38b}\\
    \text{(BC-2):}\quad & \uppartial_x \hat{c}(1, t^*) = 0, \label{eq.38c}\\
    \text{(IC):}\quad & \hat{c}(x^*, 0) = 0. \label{eq.38d}
\end{align}
\end{subequations}

\noindent
and then expressing the solution as follow:
\begin{equation}\label{eq.39}
    c^*(x^*, t^*)=\alpha^2 \int_0^t \hat{c}(x^*,\tau)\, \exp{(-\alpha^2\tau)}\text{d}\tau + \hat{c}(x^*,t^*)\, \exp{(-\alpha^2 t^*)}.
\end{equation}

Problem \eqref{eq.38} is the same as \eqref{eq.2}, with solution given by Eq.~\eqref{eq.35}. Thus, plugging this solution in \eqref{eq.39}, we get:
\begin{equation}\label{eq.40}
    c^*(x^*, t^*)= 1 - \sum_{n=1}^{\infty} \frac{2}{\lambda_n}\sin(\lambda_n x^*)\left\{\frac{\alpha^2+\lambda_n^2\,\exp{[-(\alpha^2+\lambda_n^2)t^*]}}{(\alpha^2+\lambda_n^2)}\right\},
\end{equation}

\noindent
with $\lambda_n=(n-1/2)\pi$, $n\in\mathbb{Z}^+$. The time profile is given by:
\begin{equation}\label{eq.41}
    c^*(t^*)= 1 - \sum_{n=1}^{\infty} \frac{2}{\lambda_n^2}\left\{\frac{\alpha^2+\lambda_n^2\,\exp{[-(\alpha^2+\lambda_n^2)t^*]}}{(\alpha^2+\lambda_n^2)}\right\}.
\end{equation}

\begin{figure}[t]
   \begin{center}
       \hspace{-0.1cm}\includegraphics[scale=0.425]{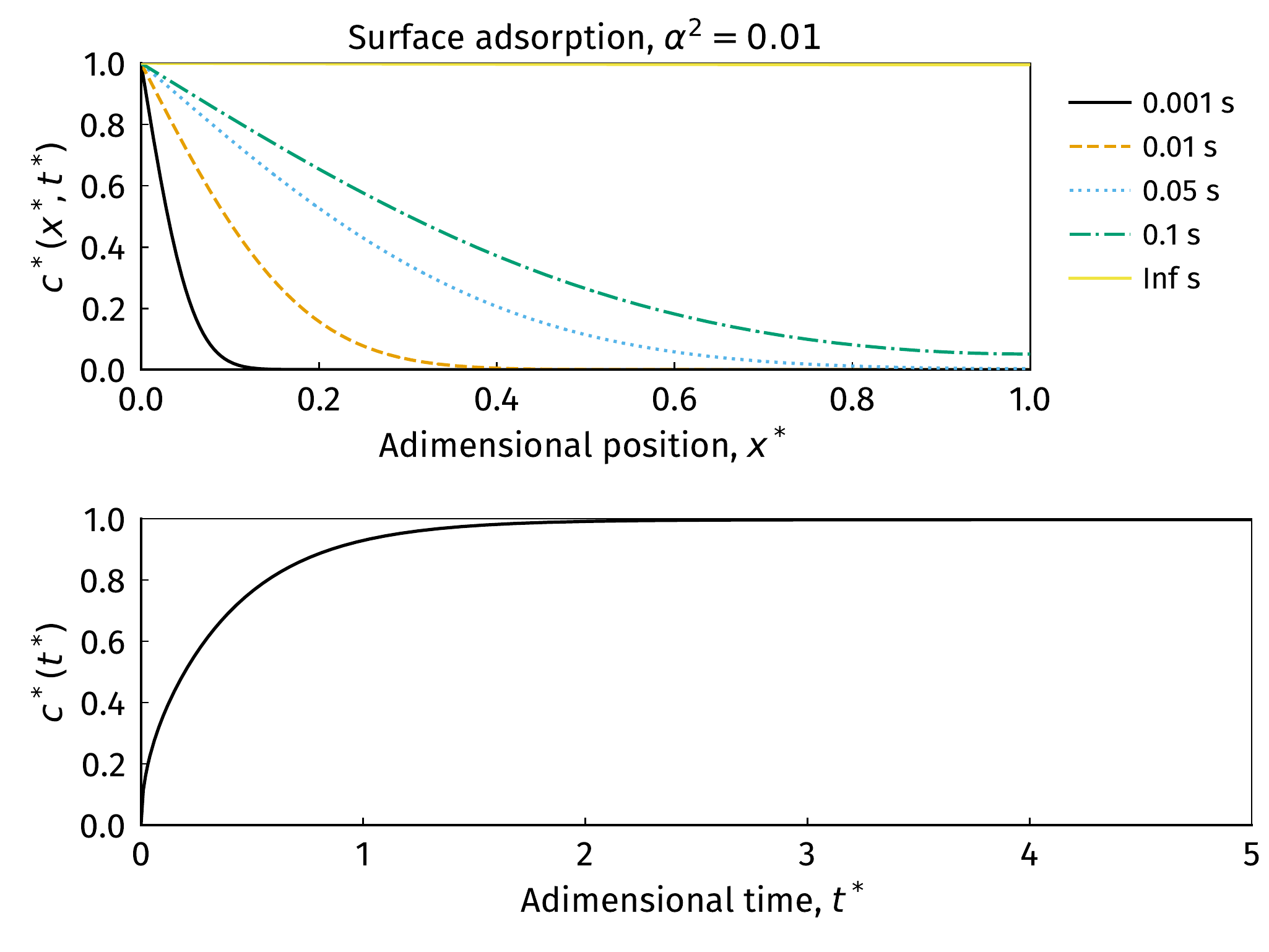}
       \caption{Concentration profiles using the separation of variables solution: (top) Eq.~\eqref{eq.40} with legends indicating different $t^*$; (bottom) Eq.~\eqref{eq.41}. For all cases, $\alpha^2=0.01$. \hfill{ }}
       \label{f.Kottke_solution_1}
   \end{center}
\end{figure}

\begin{figure}[t]
   \begin{center}
       \hspace{-0.1cm}\includegraphics[scale=0.425]{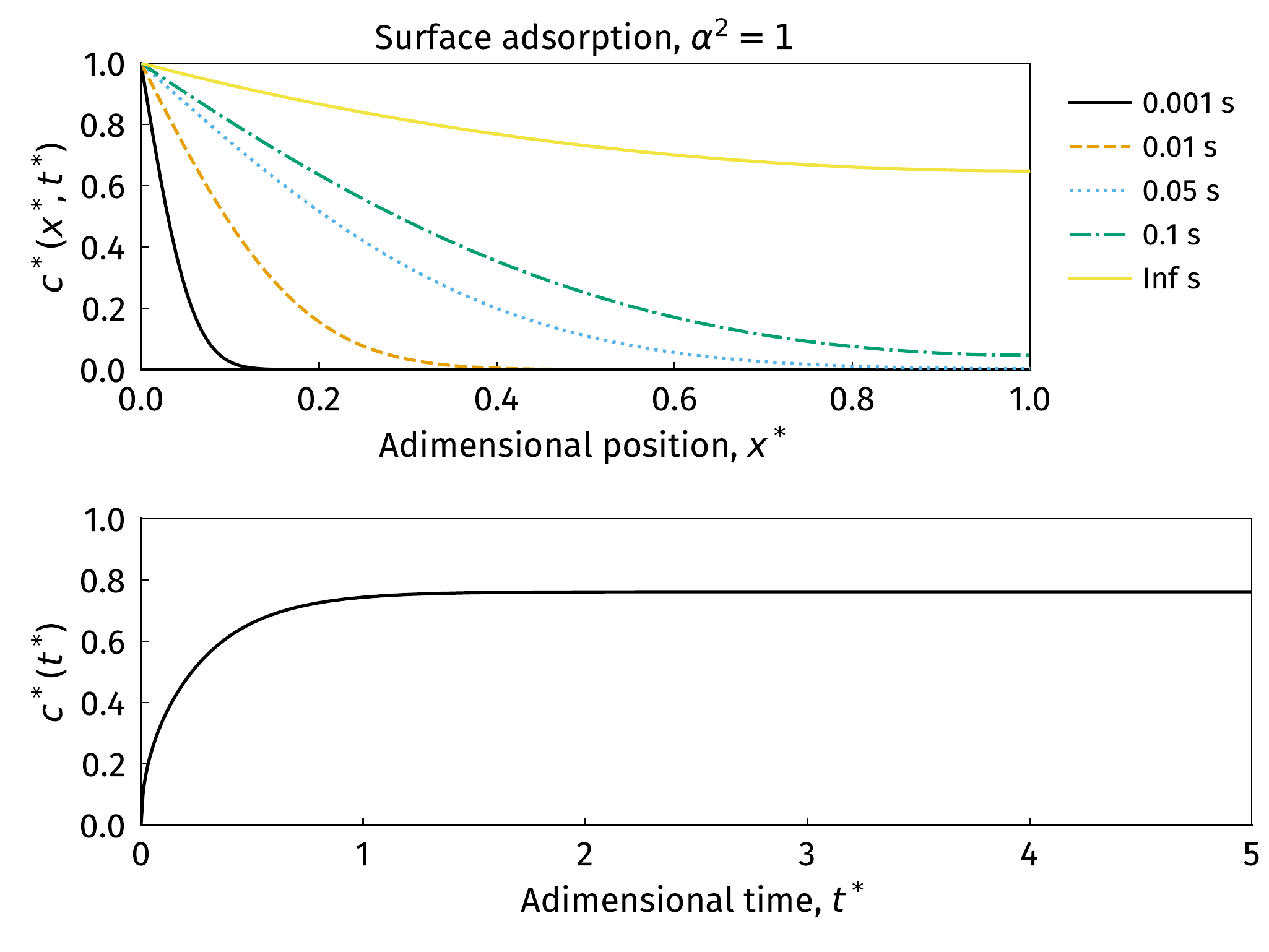}
       \caption{Concentration profiles using the separation of variables solution: (top) Eq.~\eqref{eq.40} with legends indicating different $t^*$; (bottom) Eq.~\eqref{eq.41}. For all cases, $\alpha^2=1$. \hfill{ }}
       \label{f.Kottke_solution_2}
   \end{center}
\end{figure}

Solutions of \eqref{eq.40} are presented in Figs.~\ref{f.Kottke_solution_1} and \ref{f.Kottke_solution_2} for different values of $\alpha^2$. Using Duhamel's theorem\cite{Ozisik2012} gives the same results.

\subsection{Adsorption in the wall's surface and tip of the pore}
Setting $\beta=1$ and $\gamma=1$ in \eqref{eq.1} becomes the following reaction-diffusion problem with Dirichlet and Robin boundary conditions:
\begin{subequations}\label{eq.50}
\begin{align}
    \text{(DE):}\quad & \uppartial_t c^* = \uppartial_{xx} c^* - \alpha^2 c^*, \label{eq.50a}\\
    \text{(BC-1):}\quad & c^*(0, t^*) = 1, \label{eq.50b}\\
    \text{(BC-2):}\quad & \uppartial_x c^*(1, t^*) +\alpha^2 c^*(1, t^*)=0, \label{eq.50c}\\
    \text{(IC):}\quad & c^*(x^*, 0) = 0. \label{eq.50d}
\end{align}
\end{subequations}

\noindent
We define the transformation $c^*=\exp(-\alpha^2 t^*)\,\hat{c}$, and plug it into \eqref{eq.50}, which results as follow:
\begin{subequations}\label{eq.51}
\begin{align}
    \text{(DE):}\quad & \uppartial_t \hat{c} = \uppartial_{xx} \hat{c}, \label{eq.51a}\\
    \text{(BC-1):}\quad & \hat{c}(0, t^*) = \exp(\alpha^2 t^*) = f_1(t^*), \label{eq.51b}\\
    \text{(BC-2):}\quad & \uppartial_x \hat{c}(1, t^*) +\alpha^2 \hat{c}(1, t^*) = 0 = f_2(t^*), \label{eq.51c}\\
    \text{(IC):}\quad & \hat{c}(x^*, 0) = 0 = F(x^*). \label{eq.51d}
\end{align}
\end{subequations}

\noindent
Notice that the transformation introduced not only removes the reaction term, but also replaces the Dirichlet BC-1 with a time-dependent condition, $f_1(t^*)$. For convenience, we wrote $f_2(t^*)$ as the BC-2, and $F(x^*)$ to the IC, although they are homogeneous. We solve problem~\eqref{eq.51} using the Green's function method\cite{Ozisik2012}, where the solution expressed in terms of the Green's function, $G(x^*,t^*|x^\dagger,\tau)$, is as follow:
\begin{align}
    \hat{c}(x^*, t^*) &= \int_0^1 G(x^*,t^*|x^\dagger,0)\,F(x^\dagger)\,(x^\dagger)^m\, \text{d}x^\dagger\nonumber\\
    &\,\,\, + \int_0^{t^*} \int_0^1 G(x^*,t^*|x^\dagger,\tau)\,g(x^\dagger,\tau)\,(x^\dagger)^m\,\text{d}x^\dagger\,\text{d}\tau \nonumber\\
    &\,\,\, + \sum_{i=1}^N \int_0^{t^*} x_i^m\,G(x^*,t^*|x_i,\tau)\,  f_i(\tau)\,\text{d}\tau\label{eq.51-a}
\end{align}

\noindent
where $(x^\dagger)^m$ is the Sturm-Liouville weight function with $m=0$ for rectangular spatial coordinate, $x_i$ is the $i$th boundary point of the total $N$ boundary conditions prescribed, $g(x^*,t^*)$ is the generation term \hyphen{eliminated by the transformation}, $F(x^*)$ is the initial condition, and $f_i(t^*)$ are the non-homogeneous boundary conditions functions. Notice that for a Dirichlet boundary condition, we should replace $G(x^*,t^*|x_i,\tau)$ by $\uppartial_{x^\dagger}G(x^*,t^*|x_i,\tau)$.

To find the Green's function that solves the problem, we first solve the associated homogeneous problem of~\eqref{eq.51}:
\begin{subequations}\label{eq.52}
\begin{align}
    \text{(DE):}\quad & \uppartial_t \psi = \uppartial_{xx} \psi, \label{eq.52a}\\
    \text{(BC-1):}\quad & \psi(0, t^*) = 0, \label{eq.52b}\\
    \text{(BC-2):}\quad & \uppartial_x \psi(1, t^*) +\alpha^2 \psi(1, t^*) = 0, \label{eq.52c}\\
    \text{(IC):}\quad & \psi(x^*, 0) = F(x^*). \label{eq.52d}
\end{align}
\end{subequations}

\noindent
Separation of variables, $\psi(x^*,t^*)=p(x^*)q(t^*)$, leads to the related boundary value problem of \eqref{eq.52},
\begin{subequations}\label{eq.53}
\begin{align}
    \text{(DE):}\quad & p'' + \lambda p= 0, \label{eq.53a}\\
    \text{(BC-1):}\quad & p(0) = 0, \label{eq.53b}\\
    \text{(BC-2):}\quad & p'(1) +\alpha^2 p(1) = 0, \label{eq.53c}
\end{align}
\end{subequations}

\noindent
with eigenfunctions $p_n(x^*)=\sin(\lambda_n x^*)$ and the corresponding eigenvalues $\lambda_n$, defined as the roots of the trascendental equation, $\lambda_n\cot\lambda_n+\alpha^2=0$. When $n$ tends to infinity, $\lambda_n\cot\lambda_n\sim (n-1/2)\pi$. Thus, the eigenvalues express as $\lambda_n = (n-1/2)\pi+\alpha^2$, $n\in\mathbb{Z}^+$. The difference comparing boundary value problems \eqref{eq.53} with \eqref{eq.22}, lies in the eigenvalues due to the BC-2 conditions \hyphen{Robin and Neumann} of the problem. The time-dependent solution is $q_n(t^*)=q(0)\exp(-\lambda_n^2 t^*)$, where the initial condition is obtained using the same procedure as before in Eqs.~\eqref{eq.30}-\eqref{eq.33}:
\begin{equation}\label{eq.54}
    q(0) = 2 \int_0^1 p_n(x^\dagger)\, F(x^\dagger)\, \mathrm{d}x^\dagger.
\end{equation}

\begin{figure}[t]
   \begin{center}
       \hspace{-0.1cm}\includegraphics[scale=0.425]{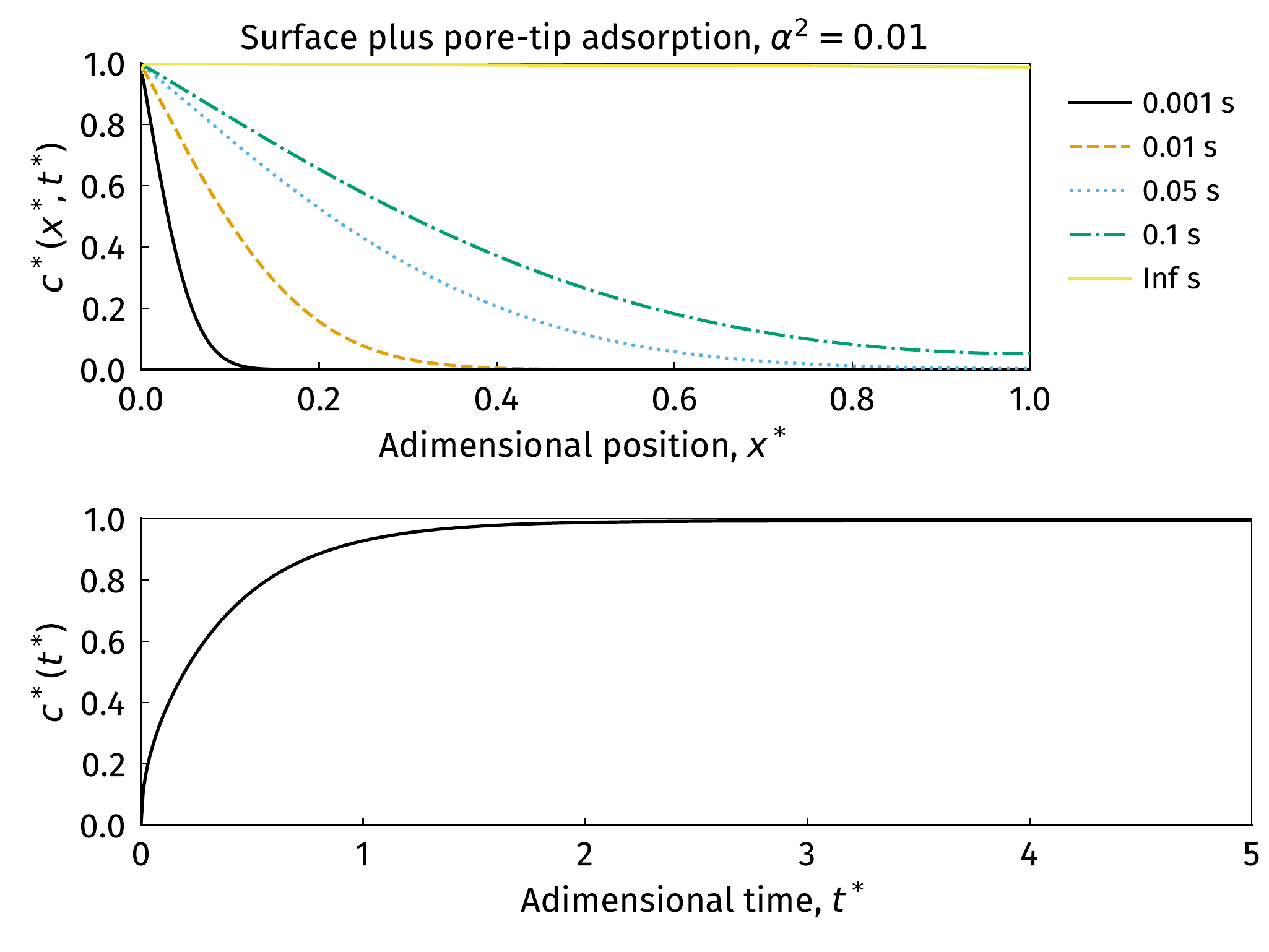}
       \caption{Concentration profiles using the separation of variables solution: (top) Eq.~\eqref{eq.51} with legends indicating different $t^*$; (bottom) Eq.~\eqref{eq.52}. For all cases, $\alpha^2=0.01$. \hfill{ }}
       \label{f.proposed_solution}
   \end{center}
\end{figure}

\noindent
The solution of \eqref{eq.52} is
  \begin{align}
    \psi(x^*,t^*) &= \sum_{n=1}^{\infty} p_n(x^*)\,q_n(t^*)\nonumber\\
    &= \sum_{n=1}^{\infty} p_n(x^*)\,q(0)\,\exp(-\lambda_n^2 t^*)\nonumber\\
    &= \int_0^1 \left[\sum_{n=1}^{\infty} 2 p_n(x^*) p_n(x^\dagger) \exp(-\lambda_n^2 t^*)\right]\, F(x^\dagger)\, \mathrm{d}x^\dagger.\label{eq.55}
\end{align}

\noindent
Last equation establishes the solution to the associated homogeneous problem. Rearranging Eq.~\eqref{eq.55} expressed in the form
\begin{equation}\label{eq.56}
    \psi(x^*,t^*) = \int_0^1 K(x^*,x^\dagger,t^*)\, F(x^\dagger)\, \mathrm{d}x^\dagger.
\end{equation}

\noindent
implies that all the terms in the solution, except the initial condition function, are lumped into a single term $K(x^*,x^\dagger,t^*)$, called the kernel of the integration. Considering the Green's function approach for the solution of the same problem, the Green's function evaluated at $\tau=0$ is equivalent to the kernel of integration: $G(x^*,t^*|x^\dagger,0)\equiv K(x^*,x^\dagger,t^*)$.
Hence, the kernel obtained by rearranging the homogeneous part of the transient problem into the form given by Eq.~\eqref{eq.56}, represents the Green's function evaluated at $\tau=0$. The full Green's function, $G(x^*,t^*|x^\dagger,\tau)$, is obtained replacing $t$ by $t-\tau$ in $G(x^*,t^*|x^\dagger,0)$\cite{Ozisik2012}. Thus, the full Green's function determined from Eq.~\eqref{eq.55} is given by
\begin{align}
    G(x^*,t^*|x^\dagger,\tau) &= \sum_{n=1}^{\infty} 2 p_n(x^*) p_n(x^\dagger) \exp[-\lambda_n^2(t^*-\tau)]\nonumber\\
    &= \sum_{n=1}^{\infty} 2 \sin(\lambda_n x^*) \sin(\lambda_n x^\dagger) \exp[-\lambda_n^2(t^*-\tau)]\label{eq.57}
\end{align}

\noindent
Now that we have calculated the Green's function, we proceed to solve problem \eqref{eq.51}, using \eqref{eq.51-a}, with $F(x^*)=0$, $g(x^*,t^*)=0$, and $f_2(t^*)=0$ in the BC-2. Also, since the BC-1 is of the first type, we replace $G$ by $\uppartial_{x^\dagger}G$:
\begin{equation}\label{eq.58}
    \uppartial_{x^\dagger}G(x^*,t^*|x_i,\tau) = \sum_{n=1}^{\infty} 2 \lambda_n \sin(\lambda_n x^*) \cos(\lambda_n x^\dagger) \exp[-\lambda_n^2(t^*-\tau)].
\end{equation}

\noindent
Then, Eq.~\eqref{eq.51-a} reduces to the following:
\begin{equation}\label{eq.59}
    \hat{c}(x^*,t^*) = \sum_{n=1}^{\infty} 2 \lambda_n \sin(\lambda_n x^*) \left[\frac{\exp(\alpha^2 t^*) - \exp(-\lambda_n^2 t^*)}{(\alpha^2+\lambda_n^2)}\right].
\end{equation}

\noindent
The final solution to problem~\eqref{eq.50} is recovered taking the anti-trasformation, $c^*=\hat{c}\,\exp(-\alpha^2 t^*)$:
\begin{equation}\label{eq.60}
    c^*(x^*,t^*) = \sum_{n=1}^{\infty} 2 \lambda_n \sin(\lambda_n x^*) \left\{\frac{1 - \exp[-t^*(\alpha^2+\lambda_n^2)]}{(\alpha^2+\lambda_n^2)}\right\}.
\end{equation}

\noindent
with $\lambda_n = (n-1/2)\pi+\alpha^2$, $n\in\mathbb{Z}^+$ and time profile as follow:
\begin{equation}\label{eq.61}
    c^*(t^*) = \sum_{n=1}^{\infty} 2 [1-\cos(\lambda_n)] \left\{\frac{1 - \exp[-t^*(\alpha^2+\lambda_n^2)]}{(\alpha^2+\lambda_n^2)}\right\}.
\end{equation}

\begin{figure}[t]
   \begin{center}
       \hspace{-0.1cm}\includegraphics[scale=0.425]{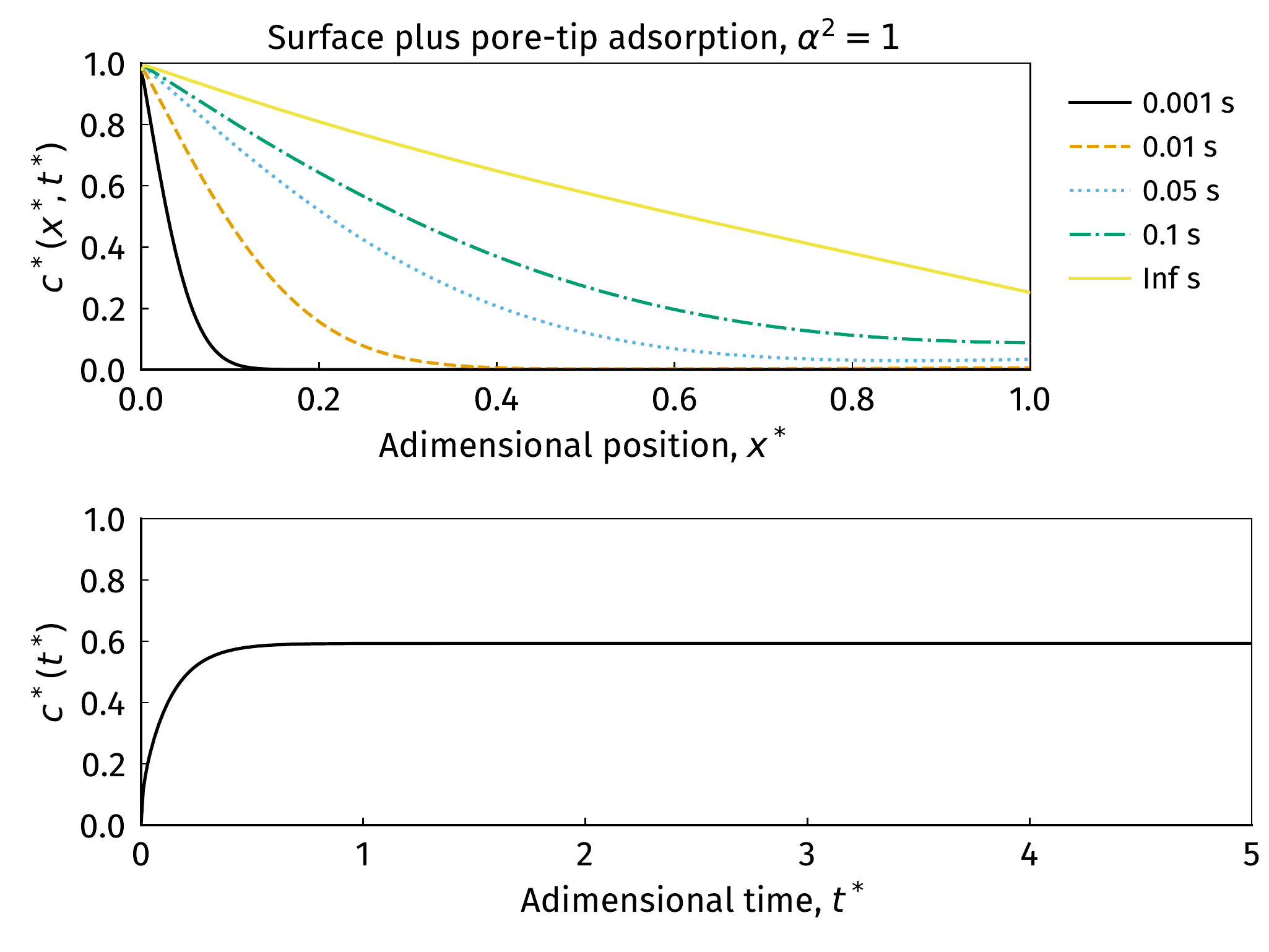}
       \caption{Concentration profiles using the separation of variables solution: (top) Eq.~\eqref{eq.51} with legends indicating different $t^*$; (bottom) Eq.~\eqref{eq.52}. For all cases, $\alpha^2=1$. \hfill{ }}
       \label{f.proposed_solution_2}
   \end{center}
\end{figure}

Figures~\ref{f.proposed_solution} and \ref{f.proposed_solution_2} show \eqref{eq.60} and \eqref{eq.61} for $\alpha^2=0.01$ and $\alpha^2=1$ for different times.

\section{Solutions to the recovery process}
We will calculate the recovery process for three cases: no adsorption from separation of variables, adsorption with zero-flux and with non-zero flux at the end of the pores. Notice that for the similarity solution, the steady-state is never reached.

\subsection{Recovery step without adsorption}
The solution of the diffusion problem without adsorption is given by Eq.~\eqref{eq.35}, where the steady-state is reached at $c^*(x^*,\infty)=1$. Then, the problem to solve is as follow:
\begin{subequations}\label{eq.63}
\begin{align}
    \text{(DE):}\quad & \uppartial_t u^* = \uppartial_{xx} u^*, \label{eq.63a}\\
    \text{(BC-1):}\quad & u^*(0, t^*) = 0, \label{eq.63b}\\
    \text{(BC-2):}\quad & \uppartial_x u^*(1, t^*) = 0, \label{eq.63c}\\
    \text{(IC):}\quad & u^*(x^*, 0) = 1. \label{eq.63d}
\end{align}
\end{subequations}

\noindent
The solution of this problem is similar to that of \eqref{eq.20}, except for the negative sign in the IC. Following the same procedure, we calculated the solution by separation of variables to be
\begin{equation}\label{eq.64}
    u^*(x^*,t^*) = \sum_{n=1}^{\infty} \frac{2}{\lambda_n} \sin(\lambda_n x^*) \exp(-\lambda_n^2 t^*).
\end{equation}

\noindent
with $\lambda_n = (n-1/2)\pi$, $n\in\mathbb{Z}^+$ and time profile
\begin{equation}\label{eq.65}
    u^*(t^*) = \sum_{n=1}^{\infty} \frac{2}{\lambda_n^2} \exp(-\lambda_n^2 t^*).
\end{equation}

\begin{figure}[t]
   \begin{center}
       \hspace{-0.1cm}\includegraphics[scale=0.425]{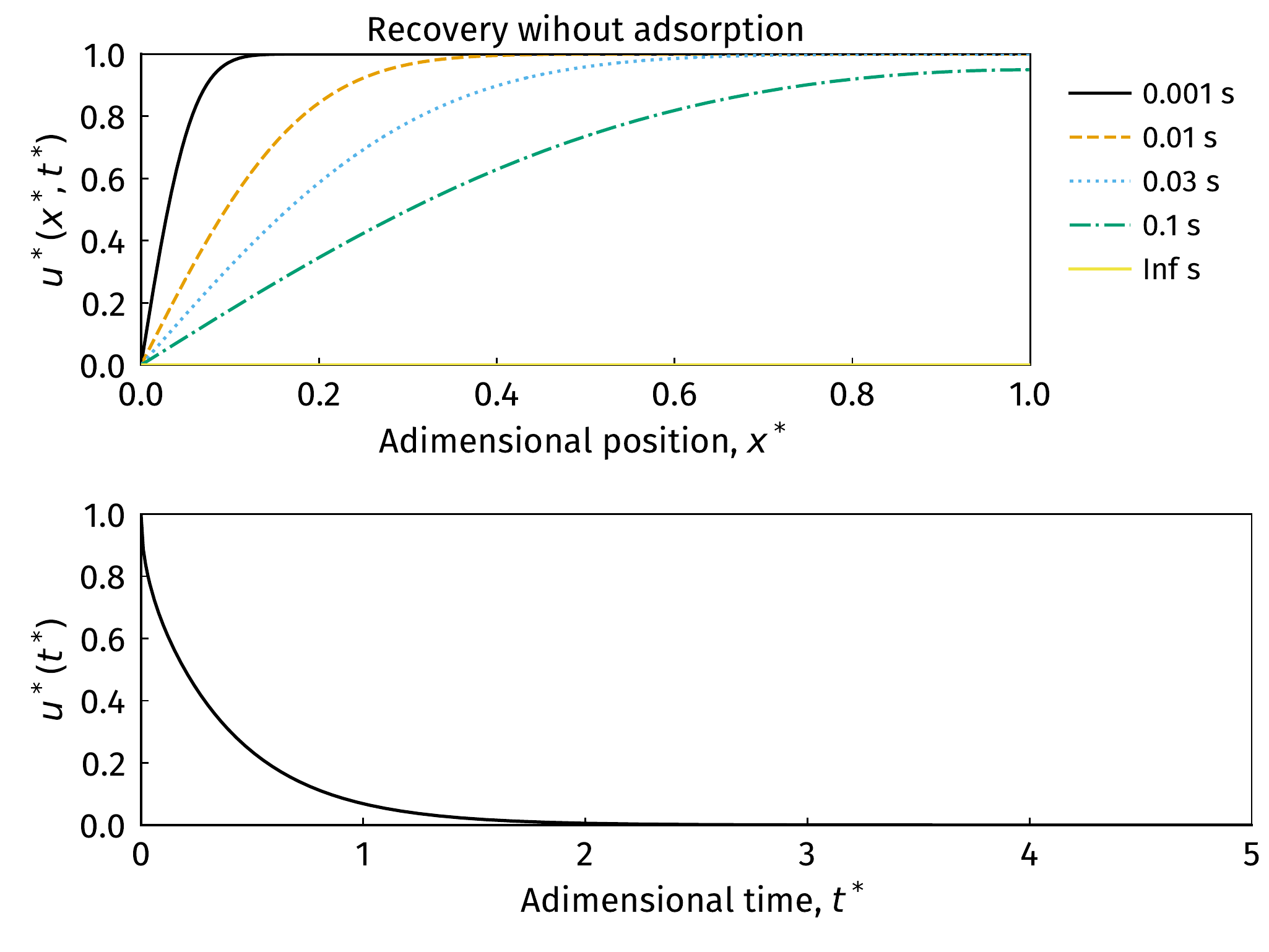}
       \caption{Concentration profiles in the recovery process without adsorption: (top) Eq.~\eqref{eq.64} with legends indicating different $t^*$; (bottom) Eq.~\eqref{eq.65}.\hfill{ }}
       \label{f.recovery_no_ads}
   \end{center}
\end{figure}

Figure~\ref{f.recovery_no_ads} shows the recovery solutions for non-adsorption.

\subsection{Recovery step with surface adsorption}

The solution of the diffusion problem is given by Eq.~\eqref{eq.40} and the recovery process by
\begin{subequations}\label{eq.66}
\begin{align}
    \text{(DE):}\quad & \uppartial_t u^* = \uppartial_{xx} u^* - \alpha^2 u^*, \label{eq.66a}\\
    \text{(BC-1):}\quad & u^*(0, t^*) = 0, \label{eq.66b}\\
    \text{(BC-2):}\quad & \uppartial_x u^*(1, t^*) = 0, \label{eq.66c}\\
    \text{(IC):}\quad & u^*(x^*, 0) = 1-\sum_{n=1}^\infty \left(\frac{2}{\lambda_n}\right) \left(\frac{\alpha^2}{\alpha^2+\lambda_n^2}\right)\sin(\lambda_n x^*). \label{eq.66d}
\end{align}
\end{subequations}

\noindent
We solve the homogeneous DE using Danckwerts' method, the problem becomes
\begin{subequations}\label{eq.67}
\begin{align}
    \text{(DE):}\quad & \uppartial_t \hat{u} = \uppartial_{xx} \hat{u}, \label{eq.67a}\\
    \text{(BC-1):}\quad & \hat{u}(0, t^*) = 0, \label{eq.67b}\\
    \text{(BC-2):}\quad & \uppartial_x \hat{u}(1, t^*) = 0, \label{eq.67c}\\
    \text{(IC):}\quad & \hat{u}(x^*, 0) = 1-\sum_{n=1}^\infty \left(\frac{2}{\lambda_n}\right) \left(\frac{\alpha^2}{\alpha^2+\lambda_n^2}\right)\sin(\lambda_n x^*). \label{eq.67d}
\end{align}
\end{subequations}

\noindent
Separation of variables, $\hat{u}(x^*, t^*)=p(x^*)q(t^*)$, leads to the boundary value problem
\begin{subequations}\label{eq.67-a}
\begin{align}
    \text{(DE):}\quad & p'' + \lambda p = 0, \label{eq.67-aa}\\
    \text{(BC-1):}\quad & p(0) = 0, \label{eq.67-ba}\\
    \text{(BC-2):}\quad & p'(1) = 0, \label{eq.67-ca}
\end{align}
\end{subequations}

\noindent
with same solution as that of \eqref{eq.22} with $p(x^*)=A_2 \sin(\lambda_n x^*)$ and $\lambda_n = (n-1/2)\pi$, $n\in\mathbb{Z}^+$. Setting $q(t^*)=q(0)\exp(-\lambda_n^2 t^*)$, we get
\begin{equation}\label{eq.68}
    \hat{u}^*(x^*,t^*) = \sum_{n=1}^{\infty} B_n \sin(\lambda_n x^*)\exp(-\lambda_n^2 t^*).
\end{equation}

\noindent
where $B_n=A_2 q(0)$. Applying the IC we calculate $B_n$ as follow:
\begin{align}
  \hat{u}^*(x^*,0) &= \sum_{n=1}^{\infty} B_n \sin(\lambda_n x^*)\nonumber\\
  \int_0^1 \hat{u}^*(x^*,0) \sin(\lambda_m x^*) \text{d}x^* &= \int_0^1 \sum_{n=1}^{\infty} B_n \sin(\lambda_n x^*) \text{d}x^*.\label{eq.69}
\end{align}

\noindent
Replacing $\hat{u}^*(x^*,0)$ by IC~\eqref{eq.67d} and using the orthogonality property of sines in Eq.~\eqref{eq.69}, we find
\begin{equation}\label{eq.70}
    B_n = \frac{2\lambda_n}{\alpha^2+\lambda_n^2}.
\end{equation}

Then,
\begin{equation}\label{eq.71}
    \hat{u}^*(x^*,t^*) = \sum_{n=1}^{\infty} 2 \left(\frac{2\lambda_n}{\alpha^2+\lambda_n^2}\right) \sin(\lambda_n x^*) \exp(-\lambda_n^2 t^*).
\end{equation}

\noindent
Combining last expression with \eqref{eq.39}, we find
\begin{equation}\label{eq.72}
    u^*(x^*,t^*) = \sum_{n=1}^{\infty} 2 \sin(\lambda_n x^*) \left\{\frac{\lambda_n[\alpha^2+\lambda_n^2\,\exp[-t^*(\alpha^2+\lambda_n^2)]]}{(\alpha^2+\lambda_n^2)^2}\right\}.
\end{equation}

\noindent
with $\lambda_n = (n-1/2)\pi$, $n\in\mathbb{Z}^+$. The time profile is given by
\begin{equation}\label{eq.73}
    u^*(x^*) = \sum_{n=1}^{\infty} 2 \left\{\frac{[\alpha^2+\lambda_n^2\,\exp[-t^*(\alpha^2+\lambda_n^2)]]}{(\alpha^2+\lambda_n^2)^2}\right\}.
\end{equation}

\begin{figure}[t]
   \begin{center}
       \hspace{-0.1cm}\includegraphics[scale=0.425]{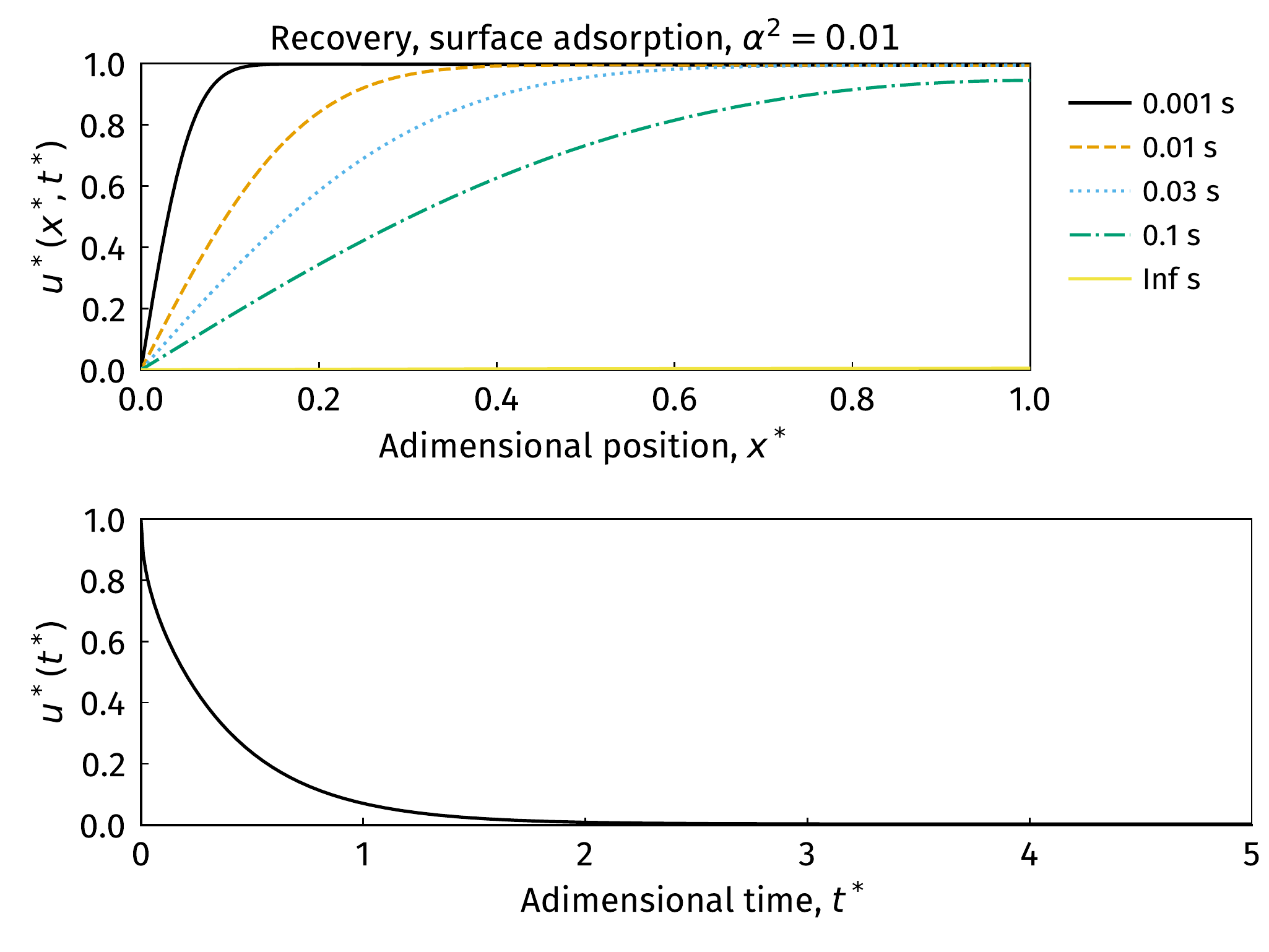}
       \caption{Concentration profiles using the separation of variables solution: (top) Eq.~\eqref{eq.72} with legends indicating different $t^*$; (bottom) Eq.~\eqref{eq.73}. For all cases, $\alpha^2=0.01$. \hfill{ }}
       \label{f.recovery_kottke}
   \end{center}
\end{figure}

\begin{figure}[t]
   \begin{center}
       \hspace{-0.1cm}\includegraphics[scale=0.425]{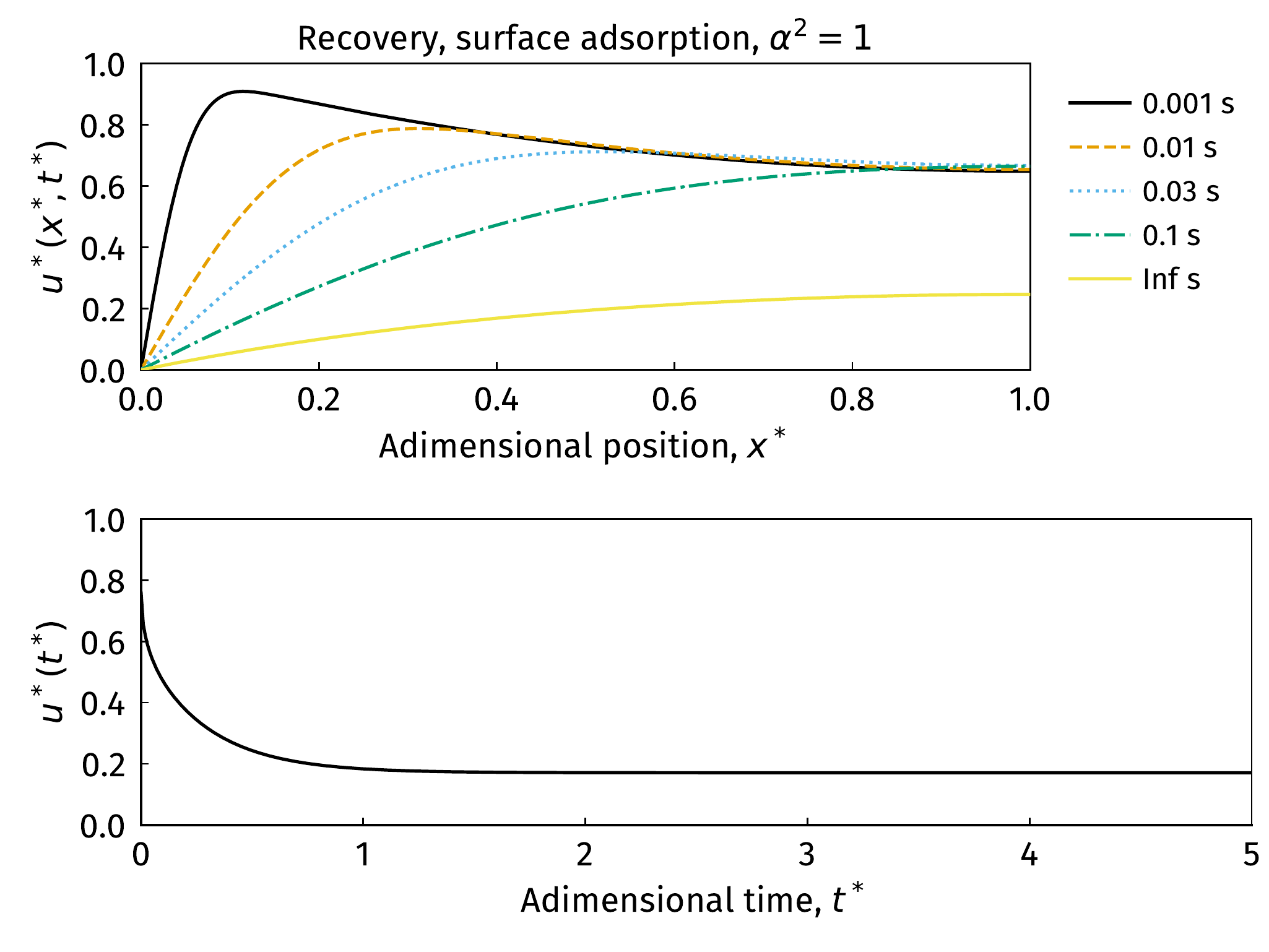}
       \caption{Concentration profiles using the separation of variables solution: (top) Eq.~\eqref{eq.72} with legends indicating different $t^*$; (bottom) Eq.~\eqref{eq.73}. For all cases, $\alpha^2=1$. \hfill{ }}
       \label{f.recovery_kottke_2}
   \end{center}
\end{figure}

Figures~\ref{f.recovery_kottke} and \ref{f.recovery_kottke_2} show the recovery solutions considering surface adsorption with zero flux at the end of the pores for two values of $\alpha$.

\subsection{Recovery step with surface and pore-end adsorption}
The problem is defined as follow:
\begin{subequations}\label{eq.74}
\begin{align}
    \text{(DE):}\quad & \uppartial_t u^* = \uppartial_{xx} u^* - \alpha^2 u^*, \label{eq.74a}\\
    \text{(BC-1):}\quad & u^*(0, t^*) = 0, \label{eq.74b}\\
    \text{(BC-2):}\quad & \uppartial_x u^*(1, t^*) + \alpha^2 u^*(1, t^*) = 0, \label{eq.74c}\\
    \text{(IC):}\quad & u^*(x^*, 0) = \sum_{n=1}^\infty \left(\frac{2\lambda_n}{\alpha^2+\lambda_n^2}\right)\sin(\lambda_n x^*). \label{eq.74d}
\end{align}
\end{subequations}

\noindent
where the IC is given by the steady-state of Eq.~\eqref{eq.60}. Using the transformation $u^*=\hat{u}\,\exp(-a^2 t^*)$ we have
\begin{subequations}\label{eq.75}
\begin{align}
    \text{(DE):}\quad & \uppartial_t \hat{u} = \uppartial_{xx} \hat{u}, \label{eq.75a}\\
    \text{(BC-1):}\quad & \hat{u}(0, t^*) = 0, \label{eq.75b}\\
    \text{(BC-2):}\quad & \uppartial_x \hat{u}(1, t^*) + \alpha^2 \hat{u}(1, t^*) = 0, \label{eq.75c}\\
    \text{(IC):}\quad & \hat{u}(x^*, 0) = \sum_{n=1}^\infty \left(\frac{2\lambda_n}{\alpha^2+\lambda_n^2}\right)\sin(\lambda_n x^*). \label{eq.75d}
\end{align}
\end{subequations}

\noindent
Separation of variables leads to the following boundary value problem:
\begin{subequations}\label{eq.76}
\begin{align}
    \text{(DE):}\quad & p''+\lambda p=0, \label{eq.76a}\\
    \text{(BC-1):}\quad & p(0) = 0, \label{eq.76b}\\
    \text{(BC-2):}\quad & p'(1) + \alpha^2 p(1) = 0, \label{eq.76c}
\end{align}
\end{subequations}

\noindent
Problem \eqref{eq.76} is the same as \eqref{eq.53}, with $p_n(x^*)=\sin(\lambda_n x^*)$ and $\lambda_n = (n-1/2)\pi+\alpha^2+\alpha^2$, $n\in\mathbb{Z}^+$. We consider a time solution of the form $q(t^*)=q(0)\,\exp(-\lambda_n^2 t^*)$, with
\begin{equation}\label{eq.77}
  q(0) = \frac{2\lambda_n}{\alpha^2+\lambda_n^2}
\end{equation}

\noindent
calculated as above. Then, the solution to \eqref{eq.75} is
\begin{equation}\label{eq.78}
   \hat{u}(x^*, t^*) = \sum_{n=1}^{\infty} \left(\frac{2\lambda_n}{\alpha^2+\lambda_n^2}\right) \sin(\lambda_n x^*) \exp(-\lambda_n^2 t^*).
\end{equation}

\noindent
Using the anti-trasformation, $u^*=\hat{u}\exp(-\alpha^2 t^*)$, we arrive at the solution of \eqref{eq.74}:
\begin{equation}\label{eq.79}
   u^*(x^*, t^*) = \sum_{n=1}^{\infty} \left(\frac{2\lambda_n}{\alpha^2+\lambda_n^2}\right) \sin(\lambda_n x^*) \exp[- t^*(\alpha^2+\lambda_n^2)]
\end{equation}

\noindent
with time profile
\begin{equation}\label{eq.80}
   u^*(t^*) = \sum_{n=1}^{\infty} 2 \left[\frac{1-\cos(\lambda_n)}{\alpha^2+\lambda_n^2}\right] \exp[- t^*(\alpha^2+\lambda_n^2)].
\end{equation}

\begin{figure}[t]
   \begin{center}
       \hspace{-0.1cm}\includegraphics[scale=0.425]{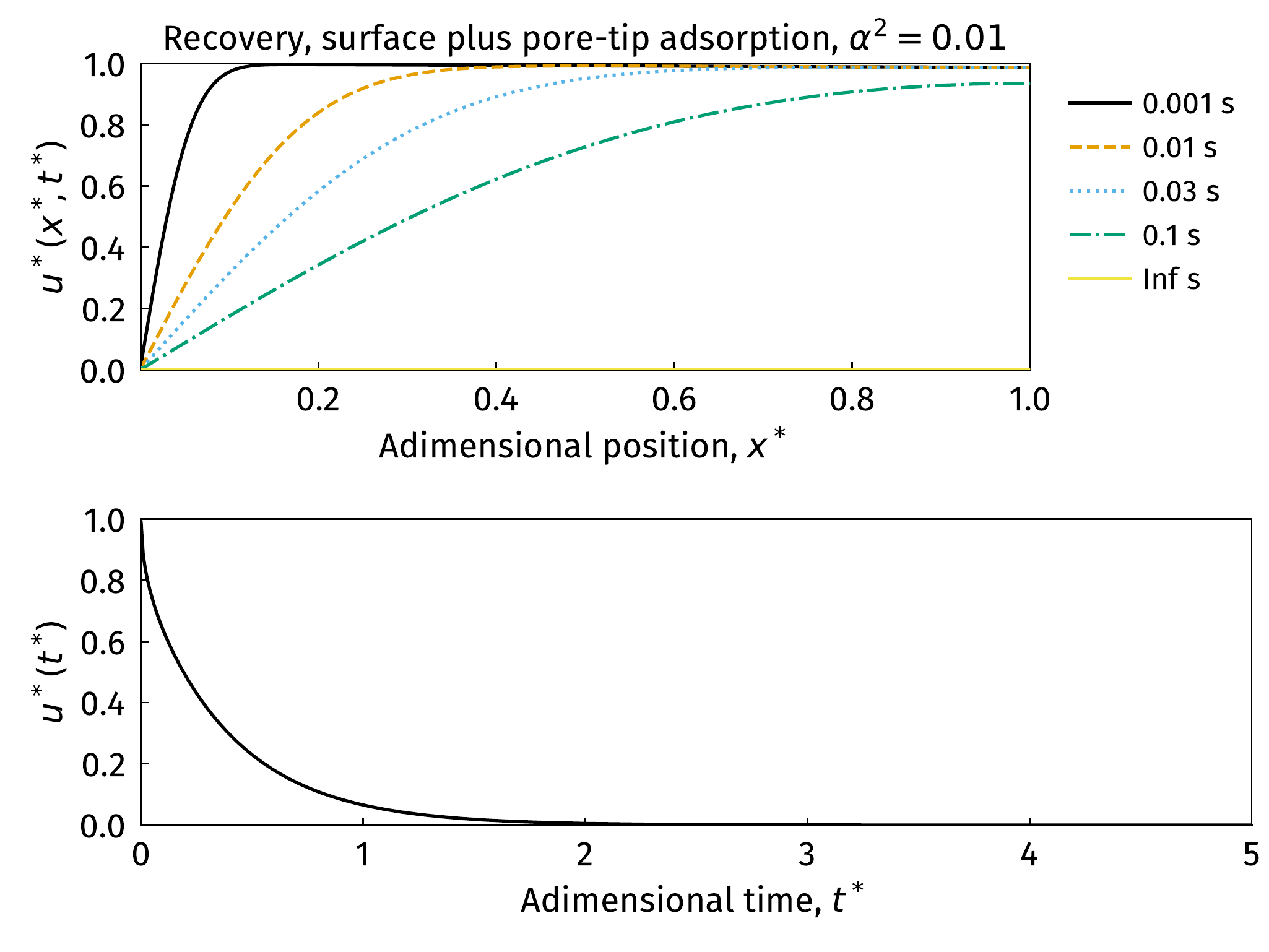}
       \caption{Concentration profiles using the separation of variables solution: (top) Eq.~\eqref{eq.79} with legends indicating different $t^*$; (bottom) Eq.~\eqref{eq.80}. For all cases, $\alpha^2=0.01$. \hfill{ }}
       \label{f.recovery_proposed}
   \end{center}
\end{figure}

\begin{figure}[t]
   \begin{center}
       \hspace{-0.1cm}\includegraphics[scale=0.425]{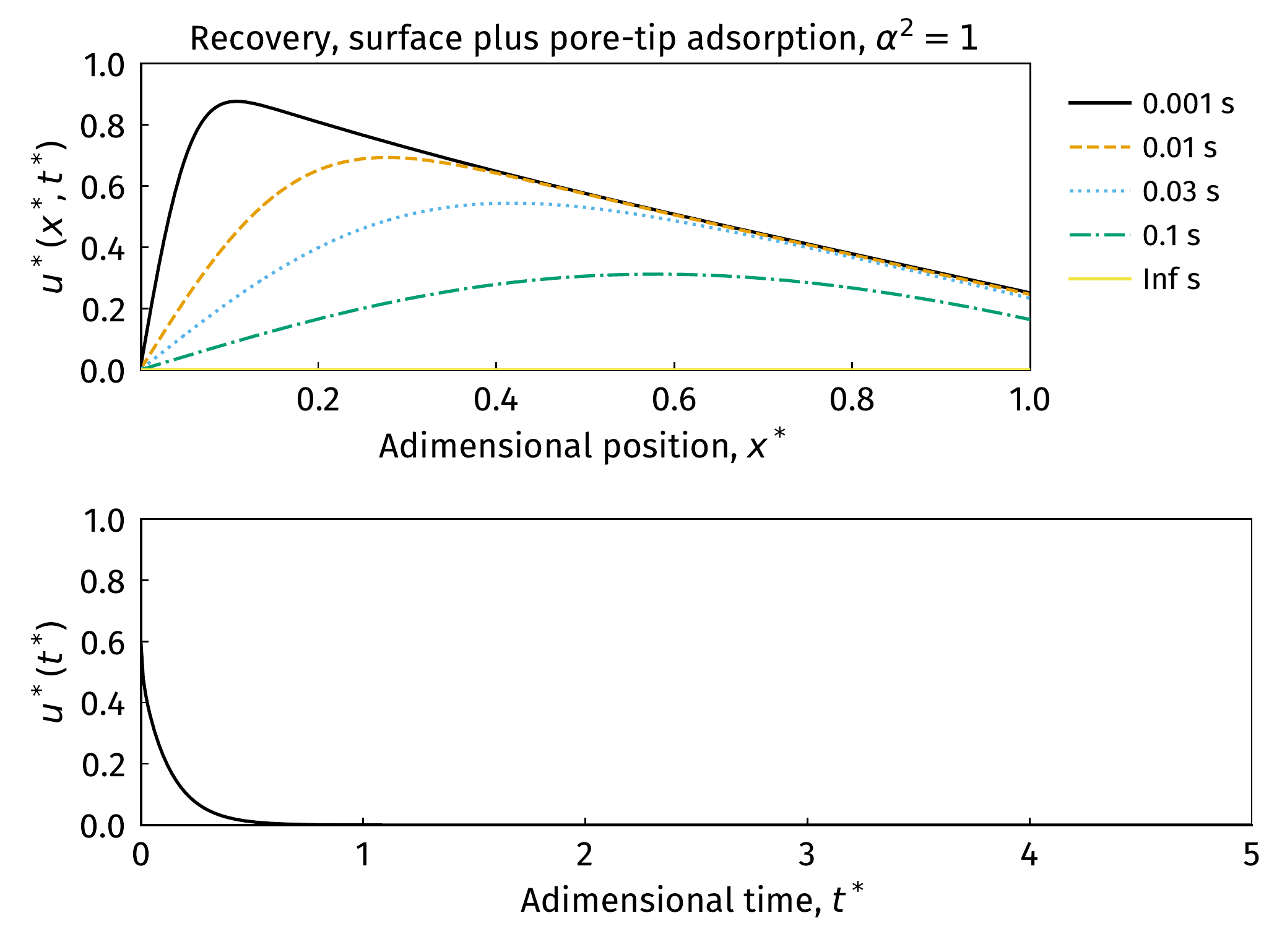}
       \caption{Concentration profiles using the separation of variables solution: (top) Eq.~\eqref{eq.79} with legends indicating different $t^*$; (bottom) Eq.~\eqref{eq.80}. For all cases, $\alpha^2=1$. \hfill{ }}
       \label{f.recovery_proposed_2}
   \end{center}
\end{figure}

Figures~\ref{f.recovery_proposed} and \ref{f.recovery_proposed_2} show the recovery solutions considering surface adsorption with zero flux at the end of the pores for two values of $\alpha$.

\linefindoc
\bibliographystyle{unsrt}

\end{document}